\documentclass[fleqn,11pt]{article}
\usepackage{hyperref}
\usepackage{amsmath, amssymb, multirow, graphicx,footnote,caption}

\usepackage{pdflscape}

\usepackage{graphicx}                    
\usepackage{float,epsfig}
\usepackage{latexsym}
\usepackage{footnote}
\usepackage{footnote}
\usepackage{amsmath}
\usepackage{amssymb}
\usepackage{amsthm}
\usepackage{url}
\usepackage{subfigure}

\makeatletter
\newcommand{\Rmnum}[1]{\expandafter\@slowromancap\romannumeral #1@}
\makeatother
\begin{document}\sloppy
\title{Asymptotic dependence modelling of the BRICS stock markets}  
\author{\footnote{${}^{1}$ Corresponding author: Tel.: +27 15 962 8135. Email address: caston.sigauke@univen.ac.za  (C. Sigauke)}\bf{${}^{*1}$Caston Sigauke, ${}^{1}$Rosinah Mukhodobwane,  ${}^{1}$Wilbert Chagwiza and ${}^{1}$Winston Garira} \\ \\
${}^{1}$Department of Mathematical and Computational Sciences, University of Venda\\ Private Bag X5050, Thohoyandou 0950, Limpopo, South Africa}

%\date{}

\maketitle

\begin{abstract}
With the use of empirical data, this paper focuses on solving financial and investment issues involving extremal dependence of ten pairwise combinations of the five BRICS (Brazil, Russia, India, China, and South Africa) stock markets. Daily closing equity indices from 5 January 2010 to 6 August 2018 are used in the study. Unlike previous literature, we use bivariate point process and conditional multivariate extreme value models to investigate the extremal dependence of the stock market returns. However, it is observed that the point process was able to model many more extreme observations or exceedances that contribute to the likelihood estimation. It gives more information than the threshold excess method of the CMEV model. This study shows varying levels of low extremal dependence structure whose outcomes are highly beneficial to investors, portfolio managers and other market participants interested in maximising investment returns and financial gains. 
\end{abstract}

\section{Introduction}
\subsection{Market linkages and extremal dependence}
\subsubsection{Stock markets linkages}
Several articles have shown that linkages between international equity markets can be described either by fundamentals or in
relation to the contagion hypothesis. The first hypothesis highlights the role of common fundamental factors (\cite{stulz1981,jawadi2013}). The fundamentals hypothesis postulates that shocks are spread through stable linkages and volatility transmission is the
same in periods of calm and crisis. The effects of contagion result when
enthusiasm for stocks in one equity market leads to enthusiasm for stocks in
other equity markets, irrespective of the evolution of market fundamentals
(\cite{jawadi2013}).

The transmission of volatility is high during market contagions as investors
discover potential changes in price using fluctuations
observed in other markets. In such a scenario, a shock stemming from one market
may present a disrupting impact on other markets, which in turn disrupts
other different markets. Occasionally, the domino effect (where the shock in
one market triggers/sets-off shock in another market(s)) ensues regardless of
the development of market fundamentals (\cite{maghyereh2012,bekaert2005}). Contagion
is the increase in the probability of crisis beyond what could be expected by
the linkages between fundamentals (\cite{fazio2007}).

Chan$-$Lau \cite{chan2004} described contagion as the probability of seeing large
return observations concurrently across different financial markets or the
co-exceedances (of extreme observations) instead of an increase in
correlations. Several monographs in the literature have documented that
financial markets' price co-movement increases significantly during the stress
period, like the Mexican crisis in 1994, the 1992-93 Exchange Rate Mechanism
crisis, the 1997 financial crises in East Asia, and 1998 in Brazil and
Russia. Chan-Lau \cite{chan2004} classified extreme (tail) events as those
returns that exceed a large threshold value (using the 95th quantile as the
cut-off point), and they used EVT methods to measure contagion as the
joint behaviour of financial returns' extremal observations
across different markets.  

The EVT approach for modelling contagion captures well the belief that small
shocks are differently transmitted across financial markets than large
shocks. The application of global extreme contagion measures to the analysis
of extreme positive and negative returns can be referred to as bull and bear
markets contagion, respectively (\cite{chan2004}).

\subsubsection{Extremal dependence}
The extremal dependence concept can be used to explain the co-movement between
markets at the tail (extreme) price or return distributions region.

Chan-Lau et al. \cite{chan2004} presented measures of extreme global contagion created
from bivariate extremal dependence measures to quantify both positive and
negative stock returns contagion at the intra- and inter-regional levels for
several emerging and mature stock markets during the past decade. The
emerging stock markets are Thailand, Taiwan Province of China, Singapore, the
Republic of South Korea, Philippines, Indonesia, Hong Kong SAR, Malaysia,
Mexico in Latin America, Brazil, Chile, and Argentina. The mature stock
markets include those of the United States, France, United Kingdom, Japan and
Germany. The authors observed that the measures of contagion using extremal
dependence and correlation approaches are not highly correlated, except for
the Latin American stock markets. Their findings suggest that results may be
misleading when correlations proxy contagion.

Using 1995-2016 stock data, risk spillovers were analysed by Warshaw \cite{warshaw2019}
across the stock markets of North America (i.e., the U.S., Canada and
Mexico), where upside (downside) risk denotes potential extreme short (long)
position losses. The dependence structure of each pair of the markets was
modelled using the generalised autoregressive score (GAS) copulas, after
which an upside and a downside Conditional Value-at-Risk (CoVaR) were
estimated. In contrast with the notion that international stock markets are
usually considered more likely to crash jointly than boom, the author
observed a symmetric conditional tail dependence for each of the paired stock
markets. The author further discovered a larger co-movement under extreme
economic conditions in all the three market pairs due to a significantly
higher symmetric conditional tail dependence following the Global Financial
Crisis (GFC). 

Using Markov Switching copulas for the periods January 1915 to February 2017, Ji et al. \cite{ji2020} analysed the risk spillovers of the equity market from the U.S.
to the other G7 nations. The authors observed that risk spillovers are asymmetric and significant and those spillovers which originated from the U.S. are stronger than those from the other G7 nations. The study further showed that upside risk spillover is weaker than downside risk spillover (\cite{warshaw2019}). In another study \cite{ji2020b} investigated the dependence and risk spillover effects from oil shocks to stock markets in BRICS countries. The study used the structural VAR model and a modeling framework that uses a time-varying copula-GARCH CoVAR approach. Results from the study indicate that there is a large risk spillover from some oil demand to the stock returns in all the BRICS countries.

The theory of co-integration analysis was used by Alagidede \cite{alagidede2008} with the
aim of analysing the market linkages among South Africa, Egypt, Nigeria and
Kenya. The findings showed that these markets do not considerably move
together, despite the economic reforms. These four
African markets (South Africa, Egypt, Nigeria and Kenya) were also modelled
by Samuel \cite{samuel2018} with daily returns data for September
2000 to August 2015 using the bivariate-threshold-excess model and point
process approach. The researchers observed that the markets displayed
asymptotic independence or (very) weak asymptotic dependence and negative
dependence.

Many authors have used the narratives of volatility spillover as a proxy for
the concept of extremal (tail) dependence. A lot of the extant articles
(e.g., \cite{hong2001,cheung1990}, among others, in the literature have focused
on risk measurement using volatility with emphasis on the effects of
volatility spillover (\cite{hong2009}). Volatility in itself is a
significant instrument in macroeconomics and finance, but in practice, it can
only effectively characterise small risks (\cite{gourieroux2001,hong2009}). In situations where extreme market movements occasionally
arise, volatility alone cannot adequately capture risk. Measures of
volatility based on distributions of asset return cannot generate accurate
market risk estimates during volatile periods (\cite{bali2000}).

An essential part of the information set required by policy makers and financial
managers is an in-depth comprehension of the direction and magnitude of
linkages and spillover effects. The knowledge of interdependence between
markets are vital to financial managers to determine diversification and
hedging of their international investment (\cite{ghini2017}).

\subsection{Reviews of studies on BRICS}
With the BRICS economies in the global spotlight, the dynamic analysis of their
markets' volatilities, risks and tail dependence are paramount to
international investors, policy-makers and all market participants who are
interested in portfolio diversifications in their stock markets. Mensi \cite{mensi2016}, examined the asymmetric linkages between the BRICS three-country risk
ratings (i.e., economic, financial and political risk) and their stock markets
from January 1995 to August 2013 with the use of a dynamic panel threshold
models. Their findings indicated asymmetry in most of the analysis, however
the signs and significance of the risk rating effects on the BRICS market
returns vary across the upper and lower regimes.

Ijumba \cite{ijumba2013} investigated the levels of interdependence and dynamic linkage
among the BRICS countries. The study employed a Vector Autoregressive (VAR),
univariate GARCH (1,1) and multivariate GARCH models. The results from VAR
showed that there is unidirectional linear dependence of Indian and Chinese
stock markets on the Brazilian market. On the other hand, the univariate GARCH
(1,1) model revealed the presence of volatility persistence in all the BRICS
markets' stock returns. China was found to be the most volatile, followed by
Russia and South Africa were the least volatile. Multivariate GARCH also
showed that there is volatility persistence among BRICS stock markets.

The industries' co-movements and pact of the member states in the BRICS
markets were investigated by \cite{lee2017} using the (BRICS) industry
weekly data from 1997 to 2013. The researchers witnessed a large increase in
the co-movements of the BRICS markets' industries effective from 2003, and
this was possibly due to the Goldman Sachs report on the BRICS economies' rapid
development. The study used GJR-GARCH and EGARCH models to determine
asymmetries in the conditional correlations of the BRICS markets’ returns.
The outcome of their work indicated signs of asymmetries on threshold and
leverage effects with a strong reaction to good news. It was further observed
that among all the sampled industries, the BRICS financial industries had the
highest co-movements.

Afuecheta \cite{afuecheta2020} contributed to the behaviour of rare events in the BRICS
stock markets. Their focus was on the extreme behaviour of the five
countries' stock markets from 1995 to 2015. The authors used five
distributions: the generalised extreme value distribution (GEVD), the
generalised logistic distribution (GLD), the generalised Pareto distribution
(GPD), the Student's $t$-one exponential parameter distribution (STED) and the
Student's $t$-two parameter Weibull distribution (STWD). The overall fit of
these distributions was compared using different criteria: log-likelihood,
the Akaike information criterion (AIC), the Bayesian information criterion
(BIC), the consistent Akaike information criterion (CAIC), the corrected
The Akaike information criterion (AICc) is the Hannan–Quinn criterion (HQC). The
outcome indicated that the GEVD was the best fit. The estimates
of value at risk, VaR$_{p}$(X) and expected shortfall, ES$_{p}$(X) from the
BRICS stock markets were computed, and it was found that Russia and Brazil
have the largest risks. The authors also modelled the tail dependence of the
BRICS economies by using various copula models, namely: Galambos,
H$\ddot{u}$sler-Reiss, Gumbel, normal and Student's $t$. models. The Gumbel
copula was the best model with the best fit.

Babu \cite{babu2015} investigated the co-movement of BRICS nations' capital
markets. This was done by investigating the short-run and long-run
integrations and linkages of BRICS countries' stock markets indices, namely,
BSE Senex, FTSE/JSE Top 40 Index, IBOVESPA, RTS Index and SSE Composite,
during the study period April 2004 - March 2014. The study employed GARCH (1,
1) model, Johannsen Co-integration test, Vector Error Correction model, and
Granger Causality test to study the stock markets linkage.  The
results that were found from the Johannsen Co-integration test revealed that all
the samples indices of BRICS stock markets were co-integrated with each
other. The study concluded that BRICS indices were engaged for a long time
relationships and only RTS Index recorded both short-run and long-run
relationships with other BRICS sample indices. The conclusion further reveals
that global investors could use the opportunity for portfolio
diversification, both under short-run and long-run periods in BRICS stock
markets.

Ji et al. \cite{ji2018} analysed the flow of information between the United States of America and BRICS equity markets. The analysis was done through the use of the DCC-MVGARCH model. The model enabled the researchers to assess the impact of specific events on information linkages across the markets. 

Bouri et al. \cite{bouri2018} examined whether implied volatility indices in some developed markets, including commodity markets, have information which can be used to predict implied volatility indices of individual BRICS countries. A robust novel method, the BGSAVR, was used in the study. Results showed a high predictive ability of the proposed model, the BGSAVR.

Also, before this study, as far as the authors know quite a few authors like \cite{mensi2016,ijumba2013,lee2017,afuecheta2020,babu2015} have
modelled the extremal dependence of the collective BRICS stock markets.
However, none of the authors has used the combined multivariate versions of
the point process models through the logistic, negative logistic,
Husler-Reiss, Bilogistic, negative bilogistic and Coles-Tawn (or Dirichlet)
models, and the CMEV model before this study to the best of the authors'
knowledge. Hence, this study robustly models and estimates the asymptotic
(extremal) dependencies in the ten pairs of the BRICS stock markets.

It is established in the literature that monetary policy shocks may influence stock market returns. The degree of the impact is driven by different contractionary impetus and the period of reserves targeting, among others. Various studies have been done, including that of \cite{gaganis2021,fullana2021}. In a study by Fullana et al. \cite{fullana2021}, the structural vector autoregressive model was used to identify a surrogate variable of monetary policy shocks. This was followed by applying a regression-based model to capture the simultaneous relationship between stock market returns and monetary policy. In their study, \cite{fullana2021} analyse whether the response of stock market returns to monetary policy shocks depends on good news or bad news and business conditions such as contraction or expansion. The results suggest that monetary policy shocks do not affect stock market returns under certain circumstances. 
	
	In another study BenSa$\ddot{i}$da et al. \cite{bensaida2018} modelled intermarket interdependences of the G7 stock markets during normal and crisis periods. The authors used a modelling framework based on the regime-switching type of copula functions. Empirical results from this study showed evidence of regime shifts in the dependence structure during turmoil periods which results in high contagion risk. In an earlier study, Boubaker and Jouini \cite{boubaker2014} carried out an empirical study on the interdependence of the equity markets of Central and South-Eastern Europe. The other study was based on the developed economies of Western Europe with the United States of America. The study used a pooled mean group modelling framework. Results from the study suggest that the stock markets are closely connected and that the impact of developed markets on emerging economies is more pronounced than the other way round.
	
	In the current study, the impact of monetary policy on stock market returns, the interdependence of stock market returns during turmoils and normal periods, including the interdependence of developed economies on emerging markets such as those of the BRICS countries, are not covered. The present paper focuses on the asymptotic (extremal) dependence of the BRICS stock market returns data from 2010 to 2018. %As such, the results should be interpreted with caution.
%In other words, the lower effectiveness of restrictive monetary policy shocks coincides with the phase of the business cycle in which the bubbles arise.	 This work finds that PM is not effective. As a result, this study will not include monetary shocks' influence on stock market returns.}
%
A summary of some previous studies on the modelling extremal dependence of stock returns of the BRICS countries, including studies from emerging and mature markets, is given in Table \ref{t1}.

%\nointerlineskip

	\begin{table}[H]
	\centering
	\caption{Summary of previous studies on extremal dependence modelling of stock returns of BRICS countries including mature markets.}\label{t1}
\scalebox{0.67}{
	\begin{tabular}{|l|l|l|l|} \hline
	{\bf Authors}  & {\bf Data}  & {\bf Models} & {\bf Main Findings} \\ \hline
	Chan-Lau et al. \cite{chan2004}  & Stock returns of some emerging and &  Extremal dependence and & Results show that the measures of \\
	                                 & some mature stock markets.   &   correlation approaches.     &  contagion are not highly correlated.\\ \hline
		Ji et al. \cite{ji2020b} & Oil demand and stock returns &  Time-varying copula-GARCH & The results show that there is a \\
	                         & of BRICS countries.           & CoVAR approach.             & large risk spillover from some  oil demand\\
	                         &                             &                             &  to the stock returns in all the BRICS countries. \\ \hline
		Samuel \cite{samuel2018}  & Daily stock returns for South Africa, & Bivariate-threshold-excess model & The markets displayed asymptotic independence \\
	                          &  Egypt, Nigeria and Kenya.            &  and point process approach    & or (very) weak asymptotic  \\
	                          &                                   &                                  &  dependence and negative dependence. \\ \hline
		Ijumba \cite{ijumba2013} & BRICS's stock returns. & Vector Autoregressive (VAR), univariate & The  Multivariate GARCH showed volatility \\
	                         &                       &  GARCH (1,1) and multivariate GARCH. &  persistence among BRICS stock markets.\\ \hline
	Afuecheta \cite{afuecheta2020} & BRICS's stock returns. & Generalised extreme value distribution, &  The results indicated that the GEVD \\
	                               &                        & generalised logistic distribution,   &  gave the best fit to the tails of the \\
	                               &                       & generalised Pareto distribution,     &  returns distributions of the   \\
	                               &                       & Student's $t$-one exponential parameter & BRICS stock markets. Using the copulas \\
	                               &                       & distribution and the Student's $t$-two &  in modelling the tail dependence, \\
	                               &                        & parameter Weibull distribution;  Galambos, & the Gumbel copula gave the best fit. \\
                                   &                        & H$\ddot{u}$sler-Reiss, Gumbel,          &                                      \\
                                   &                        & normal and Student's $t$.               &                          \\ \hline
Fullana et al. \cite{fullana2021} & Stock market returns and & Structural vector                   & Results suggest no significant   \\
                                 & monetary policy.          & autoregressive and regression models. &  monetary policy shocks  \\
                                 &                         &                                       &   on the stock market returns  \\
                                 &                         &                                       &  under certain circumstances. \\ \hline		
BenSa$\ddot{i}$da et al. \cite{bensaida2018}  & G7 stock market indices. &  Regime-switching copula models. &  Results showed evidence \\
                                             &                           &                                 & of regime shifts in the dependence \\
                                             &                          &                                  & structure during crisis periods. \\ \hline		
  	\end{tabular} }                                          
\end{table}

	\subsection{Contributions and research highlights}
	To the best of our knowledge, an in-depth analysis of asymptotic dependence modelling is not discussed in the literature. Unlike previous literature, we use bivariate point process and conditional multivariate extreme value models to investigate the extremal dependence of the stock market returns of the BRICS stock market returns. 
	
	The highlights and key findings of this study are
	\begin{itemize}
		\item The $90^{th}$ percentile is a more suitable choice in preference to the higher variance $95^{th}$ and $99^{th}$ percentiles.
		\item The pair of Brazilian IBOV and Chinese SHCOMP markets, which have a fairly strong dependence under the CMEV modelling, produced a nearly weak dependence under the point process.
		\item Bivariate point process results showed that the model best describes all the ten paired markets is the Husler-Reiss,
		with the lowest AIC value in each pair.
		\item The entire findings were consistent with the results obtained from the CMEV modelling.
		\item The only likely exception to the consistency was between the pair of Brazilian IBOV and Chinese SHCOMP markets, which has a fairly strong dependence under the CMEV modelling but produced a nearly weak dependence under the point process.
		\item Weak extremal (asymptotic) dependence between each of the seven (out of ten) paired markets from extremal dependence modelling outcomes gives beneficial risk reduction and high investment returns through international portfolio diversifications.
		\item A fairly good investment opportunity derivable from international portfolio diversifications can also be expected because the extremal dependence between the markets in these market pairs is ``fairly strong" as compared to the ``weak asymptotic" dependence.
	\end{itemize}
	
	The rest of the paper is organised as follows. Section 2 presents the materials and methods used in the study, while the empirical results are given in Section 3. A detailed discussion of the results is given in Section 4, while the conclusion is given in Section 5.
	
	%%%%%%%%%%%%%%%%%%%%%%%%%%%%%%%%%%%%%%%%%%
	\section{Materials and Methods}
	
	\subsection{Conditional multivariate extreme value modelling}
	%\subsection{The multivariate case}
	This study applies the multivariate analysis approach of \cite{heffernan2004} for the extremal dependence modelling. Before estimating the
	dependence structure, this process uses a conditional multivariate approach by
	first fitting the marginal variables with the GPD models. As the GPD model is
	used for approximating exceedances above a threshold, this dependence
	the structure is also conditioned on a variable exceeding a large enough
	threshold. To illustrate this on the BRICS stock markets, given the threshold
	exceedance of one of the markets' variables, the conditional multivariate
	the approach can describe the conditional distribution of the remaining four
	markets, with the use of a regression type model.
	
	\subsubsection{Marginal transformation}
	Before using the regression type structure for modelling dependence, the
	original data scale must be marginally transformed to standard Laplace or
	Gumbel margins. We transform to the Laplace margin %(see Theorem \ref{nbmv})
	because it simplifies the regression model's structure more than when
	transformed to the Gumbel margins (\cite{southworth2016}.
	
	Let a $p$-dimensional random variable having arbitrary marginal
	distributions be represented by $\ddot{\textbf{X}} = (X_{1}, ..., X_{p})$.
	Let an estimate of the $i^{th}$ marginal distribution function $(i = 1, ...,
	p)$ be denoted by $\hat{F}_{i}$, and let the standardised marginal
	distribution has its distribution function denoted by $G_{s}$. A transformed
	variable $\ddot{\textbf{Y}} = (Y_{1}, ..., Y_{p})$ having standardised
	marginal distributions is obtained from the original random variable
	$\ddot{\textbf{X}}$, using the probability integral transform as follows:
	
	\begin{equation}\label{inttghnnbmc}	
		\ddot{Y}_{i} ~= ~(G_{s}^{-1}(\hat{F}_{i}(X_{i}))), ~i = 1, ..., p.
	\end{equation}
	
	\subsubsection{Regression model structure}\label{LondyKumSamuel}
	Following the marginal transformation of the variables' data,
	describes the regression type structure used
	by the conditional extreme value (CEV) model for the dependence modelling.
	
	The approach used by Heffernan and Tawn is conditioned on $\ddot{\textbf{Y}}_{i}$
	being above some high threshold $u$, and the dependence of the remaining
	$\ddot{\textbf{Y}}_{-i}$ is modelled conditional on the observed value of
	$\ddot{Y}_{i}$ exceeding the threshold $u$ i.e. $\ddot{Y}_{i} > u$. The
	specific choice of $G_s$ in equation (\ref{inttghnnbmc}) dictates the form of
	the regression type model for the conditional dependence structure.
	
	\subsubsection{Laplace margins}
	\label{olwuromotiw} The Laplace distribution function is denoted by $G_{s}$
	and $\ddot{Y}$ are marginally Laplace distributed. Furthermore, on the
	condition that $\ddot{\textbf{Y}}_{i}$ variable exceeds a high enough
	threshold $u$, the model of \cite{heffernan2004} for the remaining variables
	$\ddot{\textbf{Y}}_{-i}$ is given in equation \eqref{nb1095hgvdxz}.
	
	\begin{equation}\label{nb1095hgvdxz}
		\ddot{\mathbf{Y}}_{-i} = \mathbf{\alpha}_{|i}\ddot{\mathbf{Y}}_{i} + (\ddot{Y}_{i})^{\mathbf{\beta}_{|i}}\Re_{|i},
	\end{equation}
	
	where $\Re_{|i}$ is a vector of residuals and $(p - 1)$ dimensional parameter
	vectors $\mathbf{\alpha}_{|i}$ and $\mathbf{\beta}_{|i}$ satisfying
	$(\mathbf{\alpha}_{|i}, \mathbf{\beta}_{|i}) \in [-1, 1]^{p-1} \times
	(-\infty, 1)^{p-1}$. Here, $\mathbf{\alpha}_{j|i}$, $\mathbf{\alpha}_{|i}$
	related with $\ddot{\mathbf{Y}}_{i}, (i \in {1, ..., p}, j \neq i)$, then $-1
	\leq \mathbf{\alpha}_{j|i}< 0$ and $0 < \mathbf{\alpha}_{j|i} \leq 1$
	correspond respectively to negative and positive association between
	$\ddot{\textbf{Y}}_{j}$ and $\ddot{\mathbf{Y}}_{i}$'s large values
	(\cite{southworth2016}).
	
	\subsubsection{Threshold selection}
	To select a sufficiently high threshold for the univariate risk
	modelling, a cautious trade-off between bias and variance must ensue. This
	is necessary to avoid having too high a threshold with few realizations with
	which to make inferences (\cite{ferro2003}), and which can also result in an increase
	of the parameter estimate's variance because of the reduced sample (\cite{hu2018}), or too low threshold to avoid bias where non-extreme or
	central observations are selected in place of extreme ones. In practice, the
	threshold is required to be suitably high to ensure a reliable asymptotic GPD
	approximation, hence reducing the bias (\cite{scarrott2012}). This
	study will use two threshold selection approaches described in the following
	sections.
	\subsubsection{Extreme value mixture models}
	The mixture models approach was built to provide an objective estimate of a
	suitable threshold with uncertainty quantification (\cite{scarrott2012}). The method is applied in this study because it is believed that the traditional fixed threshold approach is subjective and does not account for
	the uncertainty involved in choosing a threshold and in the resultant
	shape parameter estimates. The threshold is treated, by most mixture models,
	as a parameter that can be estimated with the use of standard inference
	schemes; hence they can (potentially) account for the related uncertainty on
	tail inferences (\cite{hu2018}).
	
	The mixture model is a combination of a bulk model under the threshold and
	GPD above the threshold. The model operates by dividing the
	distribution into two parts: the bulk and the tail. The bulk distribution
	contains high-density non-extreme observations with low information about the
	tail of the distribution. The tail fraction, on the other hand, has low
	density observations with high (asymptotic) information (\cite{scarrott2012}). Weibull, gamma and normal are some of the distributions
	used for the bulk model.
	
	Extreme value mixture models are implemented in the literature under the
	coverage of a full range of parametric, semi-parametric and non-parametric
	approaches for the bulk component (\cite{hu2018}). To obtain a
	suitable threshold selection for this study, the mixture models' approach
	will be narrowed down to the non-parametric extremal mixture models of
	MacDonald et al. \cite{macdonald2011}. This mixture model is the Kernel GPD model with
	tail modelling that follows a GPD and the bulk model under the threshold is
	the standard kernel density estimator.
	
	The non-parametric method is given preference over the usual parametric bulk
	approach because it is more robust to the bulk model than the parametric
	technique (\cite{yang2013}). Furthermore, if the population distribution is
	unknown, which is more likely the situation in financial returns
	modelling, the non-parametric extreme value mixture models will provide the
	best tail estimator (than the parametric and semi-parametric), but they add
	to the computational complexity, however, and over-fitting has to be carefully
	avoided (\cite{hu2018}). Hu and Scarrott \cite{hu2018} further indicated that
	flexible extreme value mixture models apply non-parametric density estimators
	beneath the threshold, following \cite{macdonald2011,tancredi2006}. The standard Kernel GPD model's distribution
	function is given as:
	
	\subsubsection*{Bulk model-based tail fraction approach:}
	\begin{equation}\label{iwalewa}
		F(x | X, \gamma, u, \sigma_{u}, \xi, \vartheta_{u}) = \left\{
		\begin{array}{ll}
			H(x|X, \gamma) & ~~ x\leq u,\\
			(1 - \vartheta_{u}) + \vartheta_{u} \times \ddot{G}(x|u, \sigma_{u}, \xi) & \hbox{}~~ x > u,
		\end{array}
		\right.
	\end{equation}
	
	where $\vartheta_{u} = 1 - H(u|X, \gamma)$.
	
	\subsubsection*{Parameterised tail fraction approach:}
	\begin{equation}\label{iwalewa2}
		F(x | X, \gamma, u, \sigma_{u}, \xi, \vartheta_{u}) = \left\{
		\begin{array}{ll}
			(1 - \vartheta_{u}) \frac{H(x | X, \gamma)}{H(u | X, \gamma)} & ~~ x\leq u,\\
			(1 - \vartheta_{u}) + \vartheta_{u} \times \ddot{G}(x|u, \sigma_{u}, \xi) & \hbox{}~~ x > u,
		\end{array}
		\right.
	\end{equation}
	
	where $H(x | X, \gamma)$ signifies the kernel density estimator's
	distribution function with parameter $\gamma$. The GPD's distribution
	function is $\ddot{G}(x|u, \sigma_{u}, \xi)$ and $\vartheta_{u}$ denotes the
	bulk model-based tail fraction. The $u$, $\sigma_{u}$, and $\xi$ represent
	the threshold, scale parameter and shape parameter, respectively.
	
	To estimate the tail fraction, the bulk model-based tail fraction benefits
	from borrowing information from the generally big bulk data. The main
	challenge with this bulk model, however, is that it exposes the estimation of the
	tail to the bulk model's misspecification (\cite{hu2018}). The
	parameterised tail fraction approach was introduced by \cite{macdonald2011} with an extra parameter ($\vartheta_{u}$) for the tail fraction, and
	it can reduce the effect of the misspecification of the bulk model on the
	tail estimates. Furthermore, the bulk model-based tail fraction is included
	in the parameterised tail fraction approach as a special case, where
	$\vartheta_{u} = 1 - H(u|X, \gamma)$, and it should be clear that either of
	the two specifications gives a proper density (\cite{hu2018}).
	
	\subsubsection{Estimation of parameters}
	Various techniques can be used to estimate the parameters of the GPD fitted
	to threshold exceedances. The techniques include, among others moment-based
	methods, graphical methods based on probability plots' versions, the Bayesian
	method, and the MLE. Each method has its merits and drawbacks, but the MLE is
	known to be adaptable to complex model-building with outstanding utility and
	that makes it particularly attractive (\cite{coles2001}). Further evidence in the
	literature shows that MLE gives good estimates when the shape parameter $\xi
	> -1/2$, this makes the technique more appropriate for estimation of
	financial return data with positive tail index of $\xi > 0$ (\cite{bensalah2000}).
	However, there are theoretical limitations associated with using the
	likelihood methods for generalised extreme value modelling. These limitations
	are potential difficulties known as the regularity conditions needed
	to validate the usual asymptotic properties connected with the maximum
	likelihood estimator.
	
	\subsection{Multivariate point processes}
	\subsubsection{Overview}
	Like the GPD approximation to excesses above high thresholds, the point
	process can equally be used to describe exceedances over a sufficiently high
	threshold. A region is defined above the selected threshold such that
	points in the region signify the extreme events or risks to the model. The point
	process approach incorporates other EVT models, including the $r$ largest
	order statistics, the block maxima and the threshold excess models. The
	development of these EVT models is a result of the representation of the
	point process, which forms a good reason for considering the approach (\cite{coles2001}).
	
	\subsubsection{Bivariate point process model}
	Multivariate point processes can be considered special cases of univariate
	point processes where a real-valued quantity is linked with each point event.
	A bivariate process of two types of events (e.g., type $m$ and type $n$) will
	be the case if the real-valued quantity takes only two possible values. The
	process of one of the events, say, type $m$ event alone, is known as a
	marginal process. A bivariate Poisson process is defined as a bivariate point
	process of which the marginal processes are Poisson processes. %(\cite{cox1972}).
	The Poisson limit is a reasonable approximation to a sequence of point
	processes on a suitable region (\cite{coles2001}).
	
	The point process characterization for the bivariate modelling of the
	extremal dependence can be described as follows: % stated in Theorem \eqref{OgaOGO}.
	
	Let $(x_{1},y_{1}),(x_{2},y_{2})$ ... be a sequence of realizations (or
	observed values) that are independent bivariate from a distribution having
	standard Fr$\acute{e}$chet margins, and satisfying the convergence for
	componentwise maxima.
	
	\begin{equation}\label{RichiesonofJesuOLU}
		\mbox{Pr}\{\ddot{M}^{*}_{x,n} \leq x, \ddot{M}^{*}_{y,n} \leq y\} \rightarrow G(x, y),
	\end{equation}
	
	where $\ddot{M}_{x,n}$ and $\ddot{M}_{y,n}$ are the maxima of sequences of
	two separate independent, univariate random variables ${X_{i}}$ and ${Y_{i}}$
	with standard Fr$\acute{e}$chet marginal distributions.
	
	That is,
	
	\begin{equation}\label{}
		\ddot{M}_{x,n} = \mbox{max} \{X_{i}\} ~\mbox{and} ~ M_{y,n} = \mbox{max} \{Y_{i}\},~~~ \mbox{for}~ i = 1, ..., n.
	\end{equation}
	
	Rescaled or standardised as
	
	\begin{equation}\label{}
		M^{*}_{x,n} = \mbox{max} \{X_{i}\}/n ~\mbox{and} ~ M_{y,n} = \mbox{max} \{Y_{i}\}/n,~~~ \mbox{for}~ i = 1, ..., n,
	\end{equation}
	
	where $G$ is a distribution function that is non-degenerate, and it takes the
	form
	
	\begin{equation}\label{RichiesonofJesuOLU}
		G(x, y) = \mbox{exp}\{-V(x, y)\}, ~x > 0,~ y > 0
	\end{equation}
	
	with $V(x, y)$ as specified in equations \eqref{olugbala} and
	\eqref{olugbalaAraye}, respectively (\cite{coles2001}).
	
	\begin{equation}\label{olugbala}
		V(x, y) = 2\int_{0}^{1} \mbox{max}\left(\frac{\ell}{x}, \frac{1 - \ell}{y}\right)dH(\ell).
	\end{equation}
	
	$H$ indicates a distribution function on [0, 1], and it satisfies the mean
	constraint
	\begin{equation}\label{olugbalaAraye}
		\int_{0}^{1}\ell ~dH(\ell) = 1/2.
	\end{equation}
	
	Now, let $\{N_{n}\}$ represent a sequence of point processes defined by
	
	\begin{equation}\label{RJesuOLU}
		N_{n} = \left\{\left(\frac{x_{1}}{n}, \frac{y_{1}}{n}\right), ..., \left(\frac{x_{n}}{n}, \frac{y_{n}}{n}\right)\right\}.
	\end{equation}
	
	Then, as $n \rightarrow \infty$, ${N_{n}}$ can be reasonably approximated by
	a non-homogeneous Poisson process $(N)$, as the limit distribution, on $(0,
	\infty) \times (0, \infty)$, such that
	
	\begin{equation}\label{RMinJesuOLU}
		N_{n}\xrightarrow[\text{}]{\text{d}} N,
	\end{equation}
	
	on a region of the type $K$ in equation (\ref{RiMinJesuOLUWA}), bounded from
	the origin (0, 0).
	\begin{equation}\label{RiMinJesuOLUWA}
		K = \{(0, \infty) \times (0, \infty)\} \backslash \{(0, x) \times (0, y)\}
	\end{equation}
	The intensity function of the Poisson process (or the limiting process) $N$
	is stated in equation (\ref{RMinJesuOLUWAMI}) and it indicates that the
	intensity factorises across angular and radial components, where $H$
	determines the angular spread of points in the process.
	
	\begin{equation}\label{RMinJesuOLUWAMI}
		\lambda (\hbar, \ell) = 2\frac{dH(\ell)}{\hbar^{2}}
	\end{equation}
	for
	\begin{equation}\label{RMinJesusOLUWAMI}
		\hbar = x + y ~ \mbox{and}~ \ell = \frac{x}{x + y}
	\end{equation}
	
	The choice of a sufficiently large threshold with the application of the
	bivariate point process entails the same consideration as that used by the
	threshold excess model. For both threshold models, the values of the selected
	thresholds intersect at the same points on the Cartesian $x$ and $y$ axes,
	and this can be used in comparing the outcomes of the two models.  The
	bivariate point process has an added advantage that it can be transformed to
	pseudo-polar coordinates from Cartesian, i.e., $(x, y) \rightarrow (\hbar,
	\ell)$, where the distance from the origin is measured in $\hbar$ units,
	while $\ell$ measures the angle on a scale of [0, 1]. The value of $\ell = 0$
	corresponds to $x = 0$ axis and $\ell = 1$ corresponds $y = 0$ axis.
	
	Like in the univariate case, all representations of the multivariate types
	can be obtained as special cases of the representation of the point process.
	This can be illustrated with the derivation of the componentwise block maxima's
	limit distribution as (\cite{coles2001})
	
	\begin{equation}\label{}
		\mbox{Pr}\{\ddot{M}^{*}_{x,n} \leq x, \ddot{M}^{*}_{y,n} \leq y\} = \mbox{Pr}\{N_{n}(K) = 0\} = \mbox{exp}\{-\Phi(K)\},
	\end{equation}
	where the point process as defined in equation \eqref{RJesuOLU} is denoted by
	$N_{n}$ and $K$ is the region defined in equation \eqref{RiMinJesuOLUWA}.
	Hence the limit of the point process is stated as
	
	\begin{equation}\label{RMinJesusOLUWAMI}
		\mbox{Pr}\{\ddot{M}^{*}_{x,n} \leq x, \ddot{M}^{*}_{y,n} \leq y\} \rightarrow \mbox{Pr}\{N(K) = 0\} = \mbox{exp}\{-\Phi(K)\},
	\end{equation}
	
	for the intensity measure
	
	\begin{align}
		\label{eqn:eq3234581}
		\begin{split}
			\Phi(K) &= \displaystyle\int_{K}^{}2\frac{d\hbar}{\hbar^{2}}dH(\ell)
			\\
			&= \displaystyle\int_{\ell = 0}^{1}\displaystyle\int_{\hbar = \mbox{min}\{x/\ell, y/(1 - \ell)\}}^{\infty}2\frac{d\hbar}{\hbar^{2}}dH(\ell)
			\\
			&= 2\displaystyle\int_{\ell = 0}^{1}\mbox{max}\left(\frac{\ell}{x}, \frac{1 - \ell}{y}\right)dH(\ell)
		\end{split}
	\end{align}
	The Poisson limit is a good approximation to the point process and
	convergence of the points is definite on regions bounded from the origin. For
	sufficiently large $\hbar$, convergence can be made simple if a region of the
	type $K = \{(x,y):x(n^{-1}) + y(n^{-1}) > \hbar_{\circ}\}$ is chosen, since
	the intensity measure $\Phi(K)$ is then stated as
	
	\begin{equation}\label{RMinJesuEMMANUEL}
		\Phi(K) = 2\displaystyle\int_{K}^{}2\frac{d\hbar}{\ell^{2}}dH(\ell) = 2\displaystyle\int_{\hbar = \hbar_{\circ}}^{\infty}\frac{d\hbar}{\hbar^{2}}\displaystyle\int_{\ell = 0}^{1}dH(\ell) = \frac{2}{\hbar_{\circ}},
	\end{equation}
	
	\begin{flushleft}
		and it is constant with respect to $H$ parameters (\cite{coles2001}).
	\end{flushleft}
	
	The different dependence structures faced in general datasets modelling can
	be considered using both symmetric and asymmetric models (\cite{coles1991}). Six parametric dependence models associated with the
	point process bivariate dependence modelling as indicated in the package
	``evd" (see \cite{stephenson2018}) is used in this study. The models are the
	logistic, negative logistic, Husler-Reiss, Bilogistic, negative bilogistic,
	and Coles-Tawn (or Dirichlet).
	
	The strength of asymptotic dependence in each of the six stated models are
	given in Table \ref{PPdStrDEP2}:
	
	\begin{table}[H]
		\centering
		\caption{The strength of dependence.}
		\label{PPdStrDEP2}
		\begin{tabular}{|l|l|l|}
			\hline
			Model & Independence & Dependence\\      \hline
			Logistic & $\alpha$ $\rightarrow$ 1 & $\alpha$ $\rightarrow$ 0 \\
			Negative logistic & $\alpha$ $\rightarrow$ 0 & $\alpha$ $\rightarrow$ $\infty$ \\
			Husler-Reiss & $\alpha$ $\rightarrow$ 0 & $\alpha$ $\rightarrow$ $\infty$ \\
			Bilogistic & $\alpha$ = $\beta$ $\rightarrow$ 1 & $\alpha$ = $\beta$ $\rightarrow$ 0 \\
			Negative bilogistic & $\alpha$ = $\beta$ $\rightarrow$ $\infty$ & $\alpha$ = $\beta$ $\rightarrow$ 0 \\
			Coles-Tawn (or Dirichlet) & $\alpha$ = $\beta$ $\rightarrow$ 0 & $\alpha$ = $\beta$ $\rightarrow$ $\infty$ \\
			\hline
		\end{tabular}
	\end{table}
	
	\subsection{Diagnostics: Model checking}
	To evaluate the quality of the fitted generalised Pareto model and point
	process on the threshold exceedances, suitable diagnostic model checking
	plots can be used. Classical diagnostic plots include the density plots,
	probability plots, return level plots and quantile plots. If the Poisson
	approximation of a point process fit and the GPD fit are reasonable models
	for modelling peaks or excesses above a threshold $u$, then both the
	probability and quantile plots, for example, should display points that are
	approximately linear, i.e., points that lie close to the unit diagonal.
	Significant departures from linearity will signify a failure in the validity
	of these models for the data (\cite{coles2001}).
	
	\subsection{Data description}
	The raw price data used for this study includes the daily closing equity
	indices of the Brazilian, Russian, Indian, Chinese and South African stock
	markets. The data was obtained from Thomson Reuters Datastream and is for the
	period $5^{th}$ January 2010 to $6^{th}$ August 2018 with 2126 observations.
	The data for each of the BRICS indices are recorded for 260 days per
	year, five trading days in a week. The sampling period was selected as South Africa was inducted into the group in 2010. As of 2018, these five states had a combined nominal GDP of 19.6 trillion United States dollars, about 23.2\% of the gross world product.
	
	The BRICS markets' indices are the IBOV (or Bovespa) index of Brazil Sau
	Paulo stock exchange, the IMOEX (Moscow Exchange) index of Russia, the Indian
	NIFTY (or NIFTY 50) index is the national stock exchange of India. Next is
	the SHCOMP (i.e., the Shanghai Stock Exchange Composite) index of China, and
	the JALSH (JSE Africa All Share) index of South Africa.
	
	%%%%%%%%%%%%%%%%%%%%%%%%%%%%%%%%%%%%%%%%%%
	\section{Results}
	
	\subsection{Multivariate extreme value modelling}
	Multivariate modelling context can either refer to the modelling of multiple
	random variables at various locations or a single variable at numerous
	locations, or even a single variable at multiple sites. The MEVT can be
	used to model the joint distribution of a multivariate process with
	dependence. In this study, the five BRICS financial variables are modelled in
	pairwise combinations using the CMEV model and bivariate point process.
	
	\subsection{Conditional multivariate extreme value model}
	This study applies the multivariate modelling method of \cite{heffernan2004}
	for the extremal dependence modelling by conditioning the dependence
	structure on a variable exceeding a suitably high threshold (\cite{southworth2016}). As an illustration, using the BRICS markets' variables, if the
	threshold exceedance of one of the markets' variables is given, then the
	conditional distribution of the remaining four markets can be described using
	a regression type approach as described in Section \ref{LondyKumSamuel}. The
	modelling process is carried out by first fitting the GPD to the margins,
	then the CMEV model is used for the dependence modelling. But before
	modelling the dependence, the paired variables are transformed into
	standardised Laplace margins. The modelling and analytical pattern begin with
	various multivariate exploratory data analysing plots as described in Section
	\ref{hgj1}.
	%
	%
	%\begin{figure}[H]
	%	\centering
	%	\includegraphics[width=13cm]{figures/figIndex1.pdf}
	%%	\includegraphics[width=13]{figures/figIndex2.pdf}
	%%	\includegraphics[width=13]{figures/figIndex3.pdf}
	%%	\includegraphics[width=13]{figures/figIndex4.pdf}
	%%	\includegraphics[width=13]{figures/figIndex5.pdf}
	%	\caption[Plots of the markets indicies and their returns]{Plots of the markets indicies and their returns.}
	%	\label{tsplots}
	%\end{figure}
	
\begin{figure}[H]
		\centering
		\includegraphics[width=10cm]{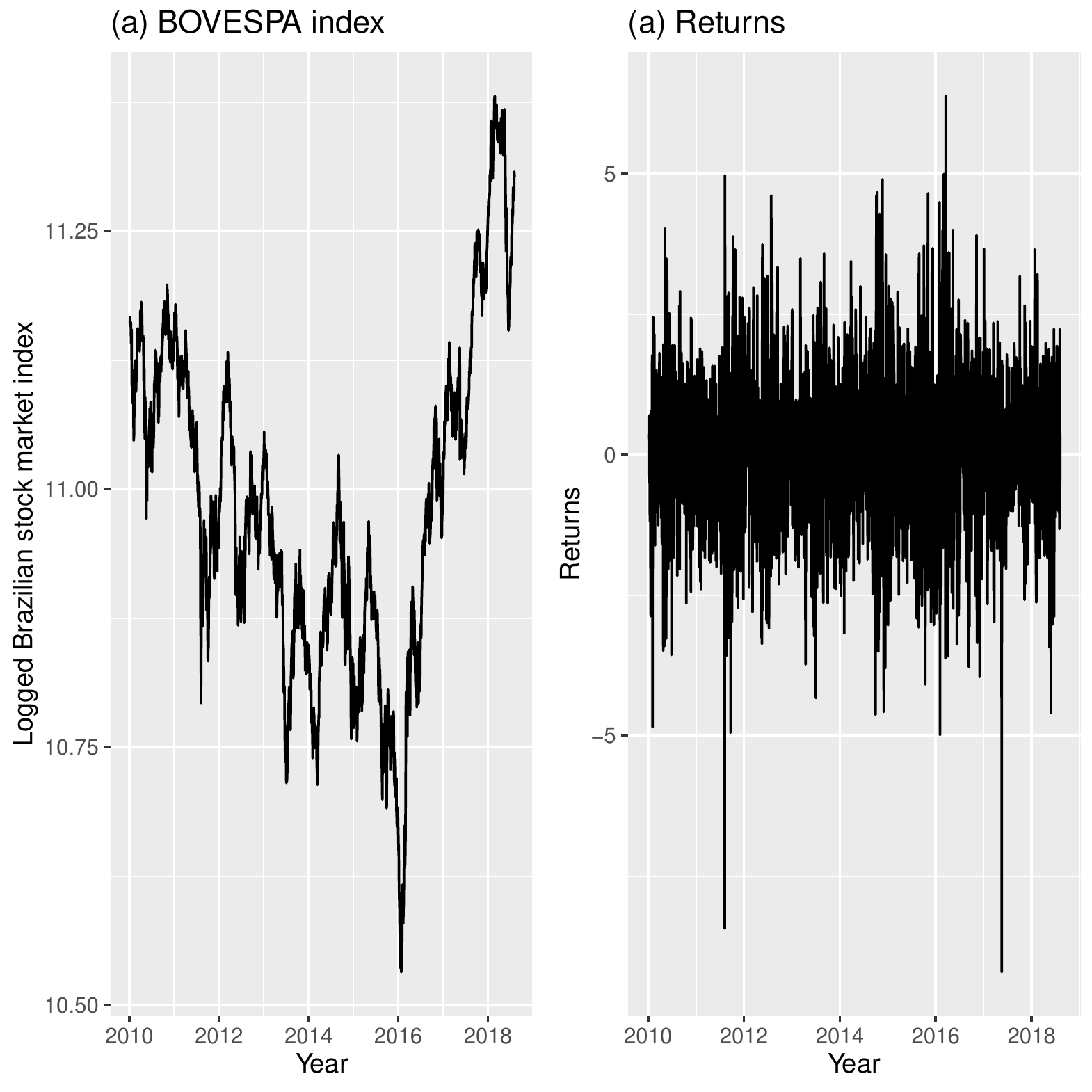}
		\caption[Left panel: Plot of Brazilian stock market index. Right panel: Plot of the returns]{Left panel: Plot of Brazilian stock market index. Right panel: Plot of the returns.}
		\label{tsplots1}
	\end{figure}

\begin{figure}[H]
		\centering	
		\includegraphics[width=10cm]{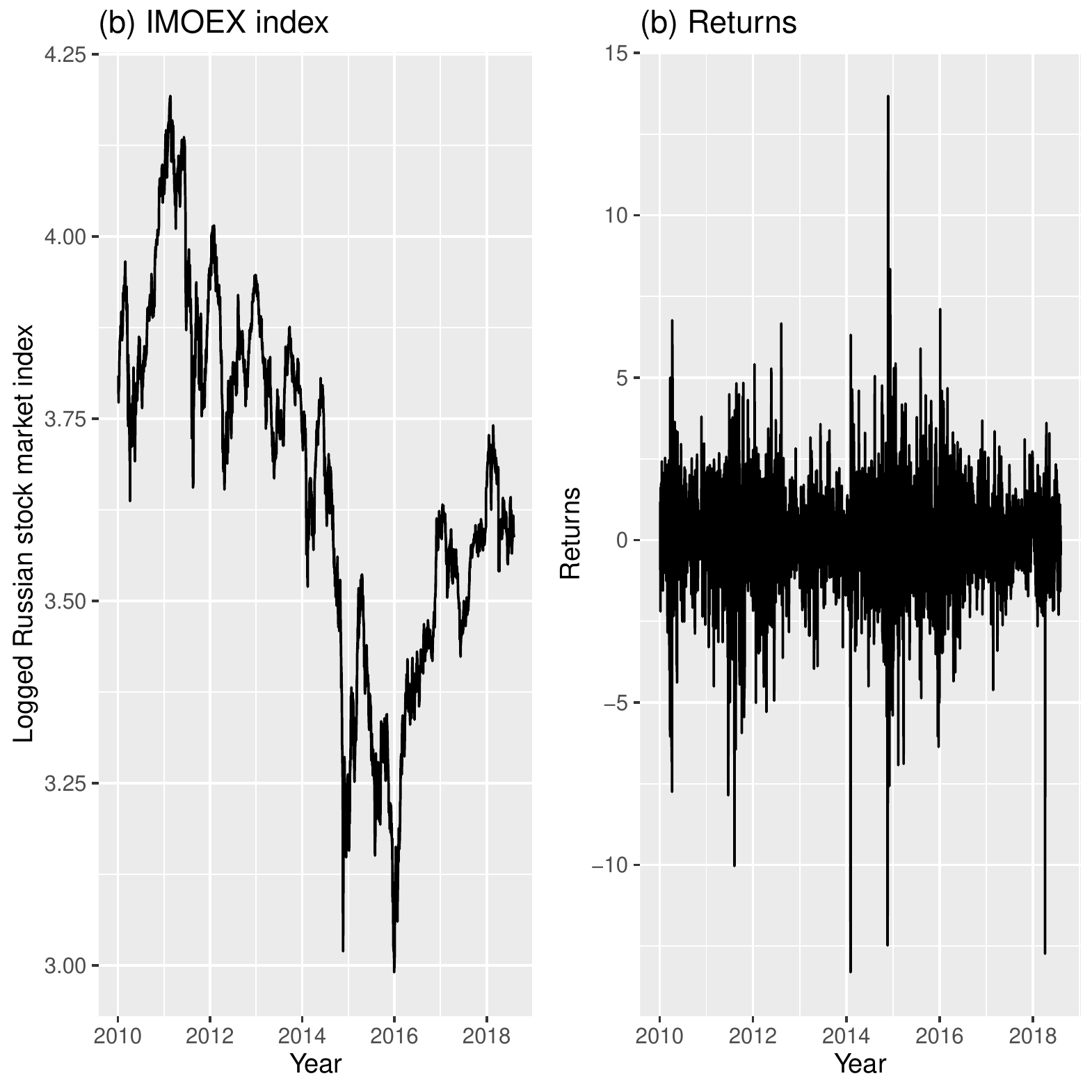}
		\caption[Left panel: Plot of Russian stock market index. Right panel: Plot of the returns]{Left panel: Plot of Russian stock market index. Right panel: Plot of the returns.}
		\label{tsplots2}
	\end{figure}

\begin{figure}[H]
		\centering
		\includegraphics[width=10cm]{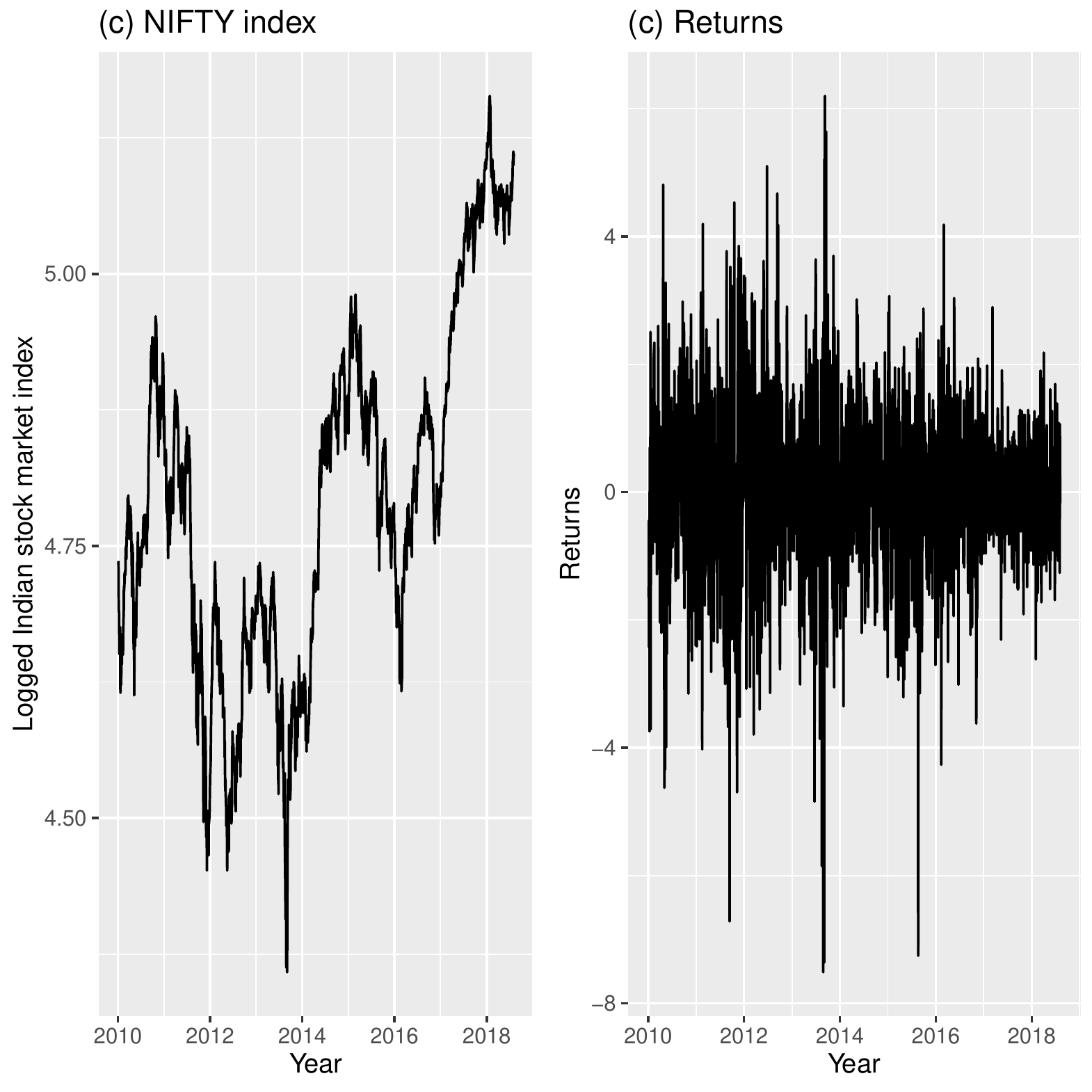}
		\caption[Left panel: Plot of the Indian stock market index. Right panel: Plot of the returns]{Left panel: Plot of the Indian stock market index. Right panel: Plot of the returns.}
		\label{tsplots3}
	\end{figure}

\begin{figure}[H]
		\centering
		\includegraphics[width=10cm]{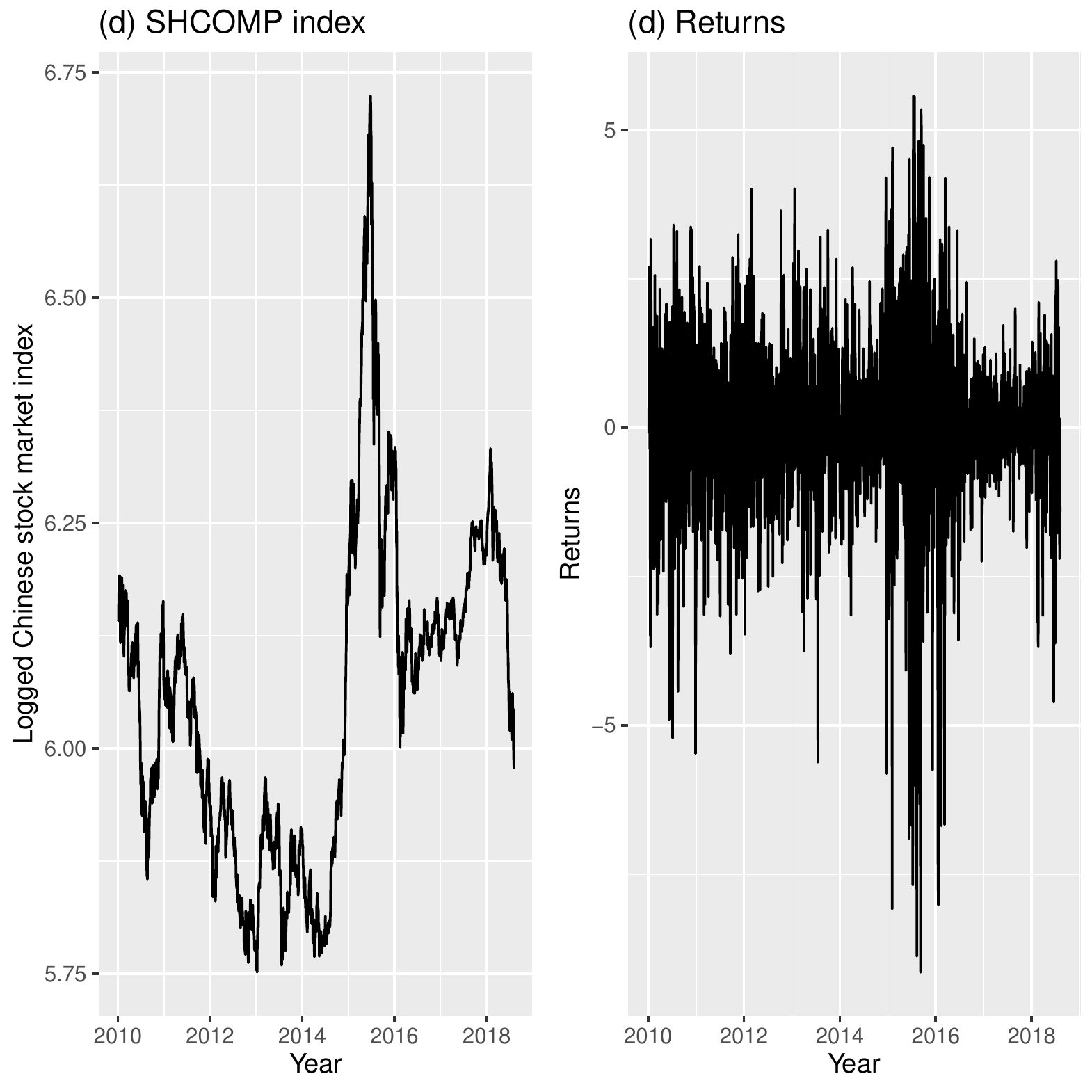}
		\caption[Left panel: Plot of the Chinese stock market index. Right panel: Plot of the returns]{Left panel: Plot of the Chinese stock market index. Right panel: Plot of the returns.}
		\label{tsplots4}
	\end{figure}

\begin{figure}[H]
		\centering
		\includegraphics[width=10cm]{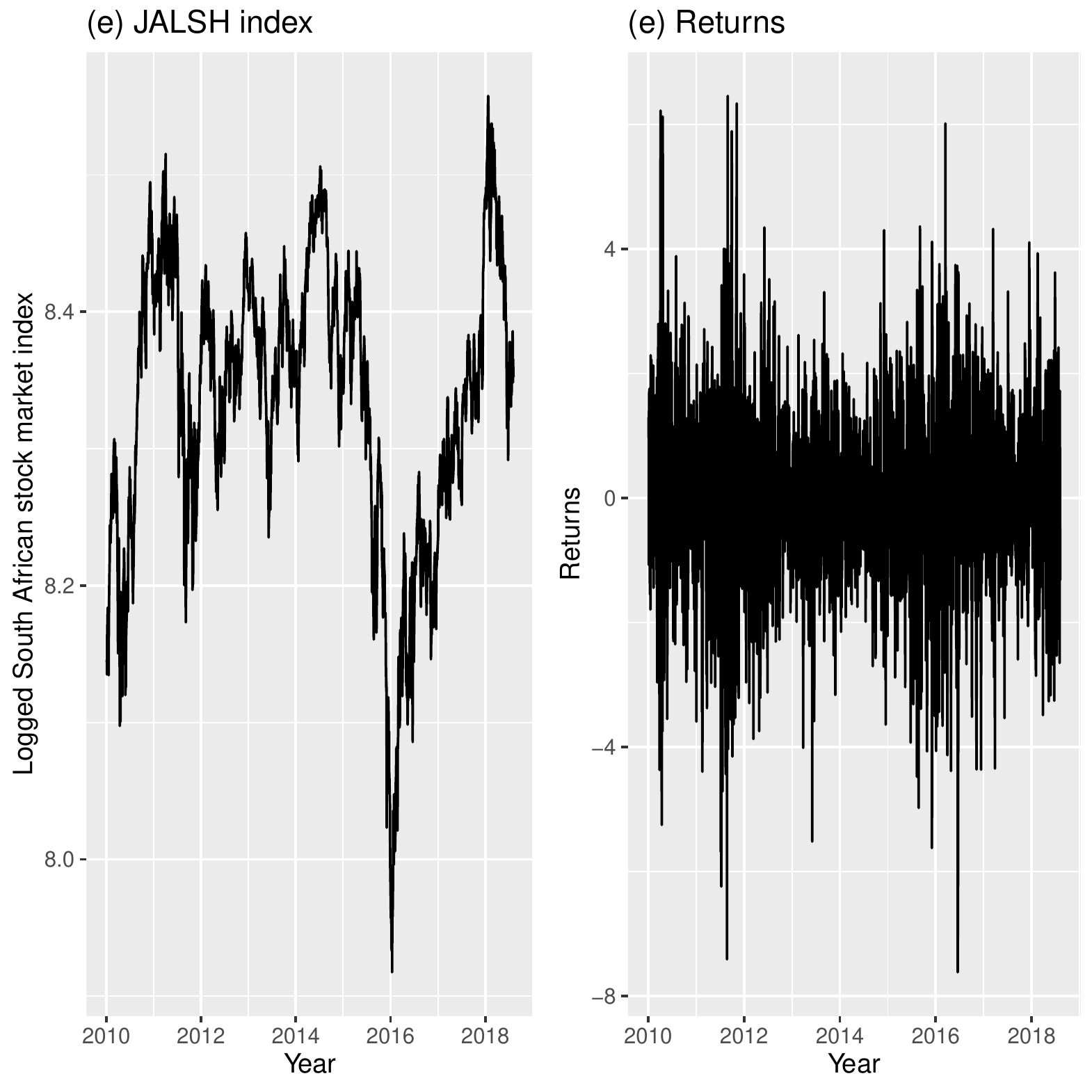}
		\caption[Left panel: Plot of the South African stock market index. Right panel: Plot of the returns]{Left panel: Plot of the South African stock market index. Right panel: Plot of the returns.}
		\label{tsplots5}
	\end{figure}

	The time series plots of the stock market indices and their returns are given in Figures \ref{tsplots1}--\ref{tsplots5}.
	
	Figure \ref{boxplotsreturns} in appendix A1 shows a comparison of the returns distributions of the BRICS stock market indices. The distributions for all the markets appear to be symmetrical with long tails.

	\subsubsection{Multivariate exploratory plots}\label{hgj1}
\begin{figure}[H]
		\centering
		\includegraphics[width=13cm]{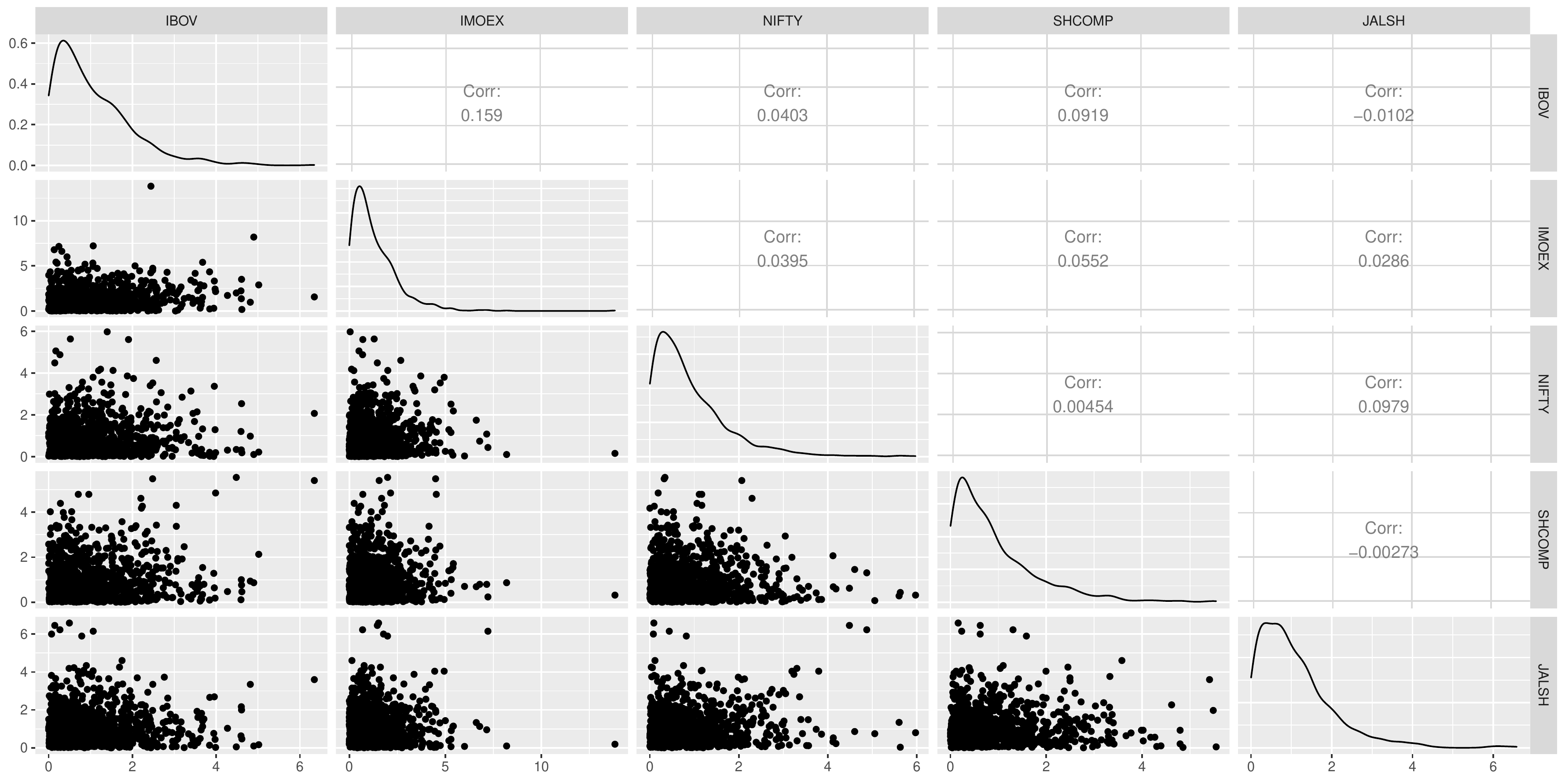}
		\caption[Pairwise scatterplot of the markets data]{Pairwise scatterplot of the markets data.}
		\label{bvnmdsfhd}
	\end{figure}

	An insightful examination using exploratory plots can be made into the
	pairwise dependence of the BRICS variables under consideration via a pairwise
	scatterplot. But it should be acknowledged that pairwise dependence between
	variables in the data body does not automatically indicate extremal
	dependence (\cite{southworth2016}). Figure \ref{bvnmdsfhd} describes the exploratory scatter plots of the five BRICS stock
	markets.
	
	The pairwise scatterplots of the BRICS residual data as shown in Figure
	\ref{bvnmdsfhd} mirror absence of extremal dependence between many of the
	paired variables, while a few market pairs like the Brazilian IBOV and
	Russian IMOEX, Brazilian IBOV and Chinese SHCOMP, Indian NIFTY and South
	African JALSH and Chinese SHCOMP and Russian IMOEX markets show some low
	levels of extremal dependence. The highest correlation value is between the
	Brazilian IBOV and Russian IMOEX markets, and it is suggesting the markets
	pair with the highest extremal dependence among the entire BRICS markets
	pairwise combinations.
	
	The quartet pairwise plots in
	the figures suggest near independence or very weak dependence across the
	board except between each pair of Brazilian IBOV and Russian IMOEX, Brazilian
	IBOV and Chinese SHCOMP, and Indian NIFTY and South African JALSH markets
	that displays weak dependence.
	
	\subsubsection{CMEV model fitting and diagnostics}
	\label{joyY} Following the preliminary analysis using the exploratory plots,
	we then proceed to the extremal dependence modelling with the application of
	the CMEV model of \cite{heffernan2004}. As stated earlier, this
	conditional multivariate modelling begins with the GPD models fitted to the
	five BRICS marginal variables, after which the dependence structure is
	estimated. In other words, the dependence component of the CMEV model also
	conditions on a variable exceeding a threshold, in the same way, as the GPD models
	exceedances above a threshold.
	
	The CMEV model is fitted to the BRICS stock dataset, conditioning on
	each of the five margins one after another. However, we need to specify an
	appropriate marginal quantile that describes the threshold above which to fit
	the marginal GPD models for each conditioning variable. To determine this
	threshold quantile, a series of candidate marginal quantiles were examined,
	and the validity of each was assessed using quartet diagnostics to ascertain
	which is the most suitable under the fitted marginal GPD model. The quartet
	diagnostics are ``probability plot", ``quantile plot," ``return level plot" and
	``histogram and density". For each market, the examined marginal quantiles
	were $70^{th}, 75^{th}, 80^{th}, 85^{th}, 90^{th}$ and $95^{th}$ percentiles, and the best of these
	quantiles, based on diagnostics appropriateness, were selected. After thorough
	examinations of the quartet diagnostic plots under the stated quantiles, it
	is observed that the most suitable threshold quantiles for the Brazilian
	IBOV, Russian IMOEX, Indian NIFTY, Chinese SHCOMP and South African JALSH
	market-variables using these diagnostics are 70th, 80th, 70th, 70th and 70th
	percentiles.
	%respectively, with the diagnostics displayed in appendix A1 in Figures
	%\ref{margQUA}, \ref{margQUA1}, \ref{margQUA2}, \ref{margQUA3} and
	%\ref{margQUA4} in that order. These diagnostic plots are the most accurate
	%when compared with the diagnostics of the remaining quantiles which were not chosen, hence
	%that informed their selections.
	
	\subsubsection{Dependence modelling and model diagnostics}
	Having obtained the marginal threshold quantiles for the GPD modelling, the
	CMEV modelling for the extremal dependence parameter estimations is carried
	out by fitting the model to the dataset in turn, conditioning on each of the
	five marginal variables. Like the marginal process, the thresholds for the
	dependence modelling is obtained by examining various candidate quantiles
	and testing their validity using diagnostic plots to know which is the most
	appropriate (\cite{southworth2016}). The examined dependence threshold
	quantiles were $70^{th}, 75^{th}, 80^{th}, 85^{th}, 90^{th}$ and $95^{th}$ percentiles. 
	
	The diagnostic plots, as shown in appendix A2 in Figures \ref{DepQuaq1}--\ref{DepQuaq5}, for the fitted CMEV
	dependence model introduced by \cite{heffernan2004} can be described
	using three diagnostic layers produced for each dependent variable. 1) The
	first layer contains residuals $\mathbf{Z}$ (from the fitted CMEV model)
	plotted against the conditioning variable's quantile, with a lowess (i.e.,
	locally weighted scatterplot smoothing) curve that shows the local mean of
	the points. A lowess is a tool that helps to see how variables relate and for
	predicting or foreseeing trends in regression analysis by creating a smooth
	line through a scatter plot. 2) The second layer contains the plots of the
	absolute value of $\mathbf{Z}$-mean($\mathbf{Z}$), where the lowess curve
	shows the local mean of the points. 3) The third layer displays the plots
	showing the original data that is not transformed and the quantiles of the
	CMEV model fit.
	
	As a condition for good fitness, the plots in layers 1 and 2 are meant to
	display a lowess curve (or scatterplot smoother) that is more or less
	horizontal (\cite{southworth2016,southworth2020}). Or, as stated
	by \cite{southworth2020}, any trend in the scatter or location of the
	variables with the conditioning variable violates the assumption of the model
	that the residuals $\mathbf{Z}$ are independent of the conditioning variable.
	Hence, the straighter (i.e., no trend) the lowess curve or scatterplot
	smoother is, the better the fit.  Furthermore, in layer 3, a model that is
	well fitted is expected to have a satisfactory agreement between the fitted
	quantile and the scatter plot (i.e., the raw data distribution) (\cite{southworth2016,southworth2020}).
	
	From the dependence, diagnostic plots in Figures \ref{DepQuaq1}--
	\ref{DepQuaq5}, it is observed that the parameter estimates of the
	the conditional multivariate extreme value model is most accurate where the
	lowess curves are smoothest at the $70^{th}$ percentile for all the BRICS stock
	markets. That is, these diagnostic plots support the selection of the $70^{th}$
	percentile dependence quantile. Moreover, at this chosen quantile, the raw
	data distribution is approximately in good agreement with the fitted
	quantiles (i.e., the solid vertical line) and the scatter plots (the dotted
	lines) as displayed in layer 3 of each market's diagnostic plots. It is
	necessary to know that different threshold quantiles can eventually be used
	for the marginal and dependence modelling (\cite{southworth2016}), since
	the choice of a suitable threshold for each of the modelling is strictly
	based on the appropriateness of the diagnostic assessment plots. This is
	observed in the Russian IMOEX market, where the $80^{th}$ percentile was obtained as
	the marginal quantile while the $70^{th}$ percentile was used as the dependence
	quantile.
	
	\subsubsection{Extremal dependence results of the CMEV model}
	Table \ref{LondyMyWife} shows the estimated parameters of the BRICS stock
	markets' dependence structure conditioning on each of the five margins, one
	after the other. That is when the threshold excesses one of the markets
	are given, the conditional distribution of the remaining four market
	variables are described. Here, we described the dependence between market
	pairs by a pair of parameters ``$A$" and ``$B$," with more attention focused
	on the former, such that the values of ``$A$" near $1$ or $-1$ indicate
	strong positive or negative extremal dependence (\cite{southworth2016}).
	
	\begin{table}[H]
		\centering
		\caption{Dependence structure parameter estimates of the CMEV model.}\label{LondyMyWife}
		\begin{tabular}{|l|l|l|l|l|l|}
			\hline
			& Dependence  & IMOEX & NIFTY & SHCOMP & JALSH\\
			& parameters &  &  &  &\\                           \hline
			Conditioning on:   & $a$ & 0.3888 & -0.0194 & 0.3157 & -0.0408\\
			IBOV &  &  &  &  &\\                           \hline
			& $b$ & 0.1019 & 0.0257 & 0.2452 & 0.0633\\
			&  &  &  & & \\                                     \hline
			& Dependence  & IBOV  & NIFTY & SHCOMP   & JALSH\\
			& parameters &  &  &  &\\                          \hline
			Conditioning on:   & $a$ & 0.1655 & 0.0151 & 0.0314 & 0.0184\\
			IMOEX &  &  &  &  &\\                             \hline
			& $b$ & 0.1386 &0.0669 & -0.1024 & 0.1650\\
			&  &  &  & & \\                                     \hline
			& Dependence  & IBOV  & IMOEX & SHCOMP & JALSH\\
			& parameters &  &  &  &\\                         \hline
			Conditioning on:   & $a$ & 0.1116 & 0.0059 & -0.1110 & 0.2531\\
			NIFTY &  &  &  & & \\                              \hline
			& $b$ & 0.0066 & 0.0140 & -0.1518 & 0.2034\\
			&  &  &  &  &\\                                     \hline
			& Dependence  & IBOV  & IMOEX & NIFTY & JALSH\\
			& parameters &  &  & & \\                            \hline
			Conditioning on:   & $a$ & 0.3159 & 0.0717 & -0.0957 & -0.1235\\
			SHCOMP &  &  &  & & \\                               \hline
			& $b$ & 0.2081 & -0.0055 & -0.0352 & 0.1015\\
			&  &  &  &  &\\                                     \hline
			& Dependence  & IBOV  & IMOEX & NIFTY & SHCOMP\\
			& parameters &  &  & & \\                            \hline
			Conditioning on:   & $a$ & -0.0606 & 0.1018 & 0.2011 & 0.0311\\
			JALSH &  &  &  & & \\                               \hline
			& $b$ & 0.0642 & -0.0138 & 0.2315 & 0.1255\\
			\hline
		\end{tabular}
	\end{table}
	
	The dependence parameter estimates of the CMEV modelling generates
	the following results as shown in Table \ref{LondyMyWife}.
	\begin{enumerate}
		\item Conditioning on Brazilian IBOV market: From the table, it is
		clearly shown that the Russian IMOEX and Chinese SHCOMP markets have
		fairly strong positive extremal dependence on large values of the
		Brazilian IBOV market, with the Russian IMOEX market having stronger
		dependence than the Chinese SHCOMP market on the Brazilian IBOV
		market. The Indian NIFTY and South African JALSH markets, on the other
		hand, have a very weak negative extremal dependence on the
		conditioning the Brazilian IBOV market.
		\item Conditioning on Russian IMOEX market: The Brazilian IBOV, Indian
		NIFTY, Chinese SHCOMP and South African JALSH markets have a
		relatively weak positive extremal dependence on the Russian IMOEX
		market, with the strongest of this weak dependence being between
		the Russian IMOEX and Brazilian IBOV markets.
		\item Conditioning on Indian NIFTY market: Here, it is observed that the
		Brazilian IBOV and Russian IMOEX markets have varying levels of weak
		positive extremal dependencies on the Indian NIFTY market, while the
		asymptotic dependence between the Indian NIFTY and South African
		JALSH markets are moderately strong. The Chinese SHCOMP market, however
		has a weak negative dependence on the Indian NIFTY market.
		\item Conditioning on Chinese SHCOMP market: The values of this
		dependence parameter shows that the Brazilian IBOV is the most
		(fairly) strongly positively dependent on large values of the Chinese
		SHCOMP market, while the Russian IMOEX, Indian NIFTY and South
		African JALSH markets have only weak extremal dependence on the
		Chinese SHCOMP market. More specifically, the Indian NIFTY and South
		African JALSH markets have weak negative levels of dependence while
		the Russian IMOEX market has a weak positive dependence on the
		Chinese SHCOMP market.
		\item Conditioning on South African JALSH market: The values of the
		dependence parameter estimates show that the Russian IMOEX, Indian
		NIFTY and Chinese SHCOMP markets all have different weak positive
		extremal dependencies on the South African JALSH market, the strongest
		of these is between the South African JALSH and Indian NIFTY markets.
		The Brazilian IBOV market has a weak negative extremal dependence on
		the South African JALSH market.
	\end{enumerate}
	
	\subsubsection{Prediction under the CMEV model}
	The fitted CMEV model can be interpreted well through variables
	prediction given extreme values of a conditioning variable (\cite{southworth2016}). With the use of ``importance sampling," prediction can be made by
	estimating quantiles or probabilities of threshold exceedances (as shown in
	Table \ref{esK1nmxc}) for the fitted CMEV model given the conditioning
	variable above the threshold for extrapolation (\cite{southworth2020}).
	
\begin{figure}[H]
		\centering
		\includegraphics[width=10.5 cm]{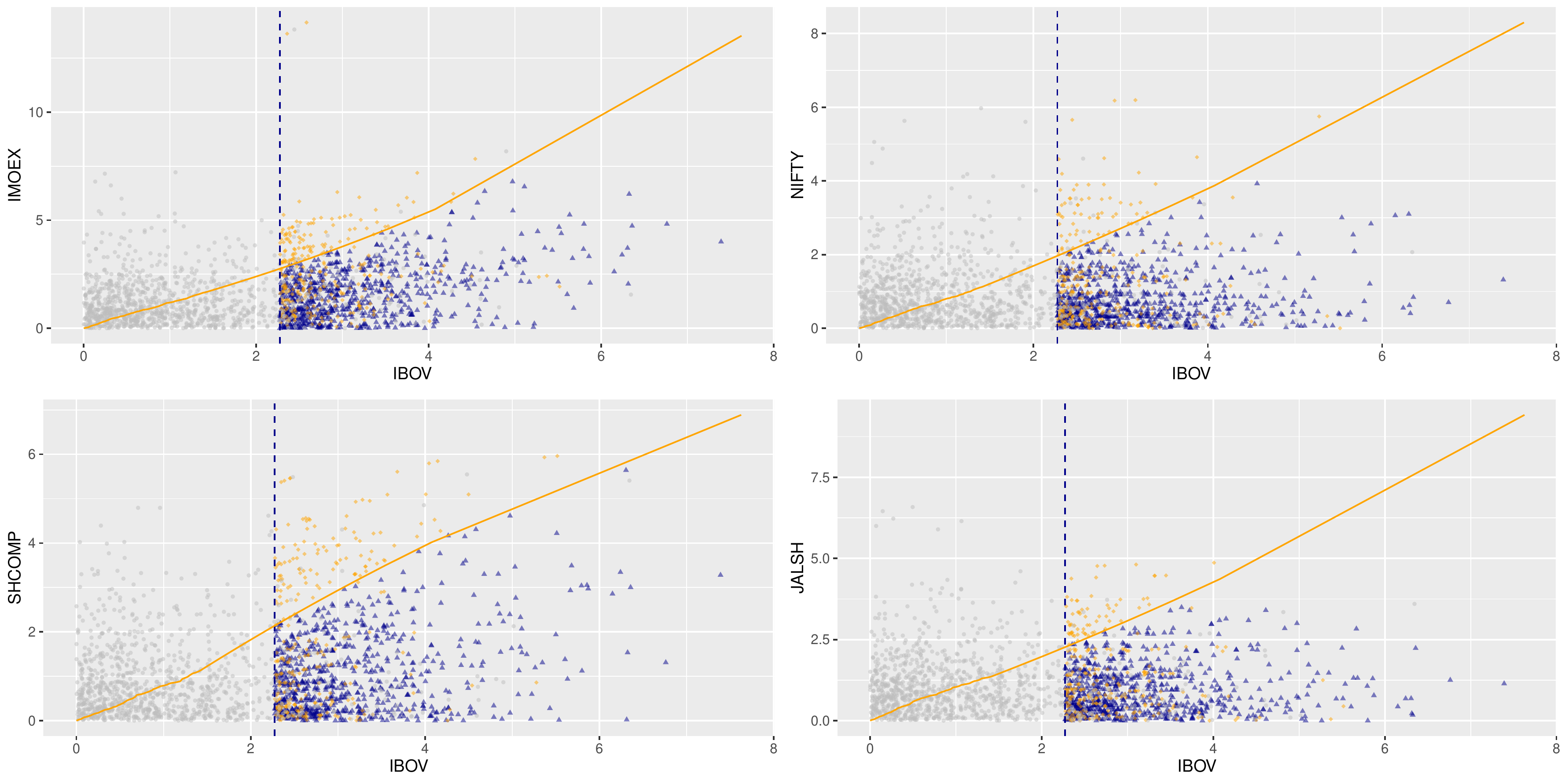}
		\caption[Prediction plot conditioning on IBOV variable]{Prediction plot conditioning on IBOV being above its $90^{th}$ percentile.}
		\label{bIBOVprdtng}
		%\end{figure}
		\begin{flushleft}
		\end{flushleft}
		%\begin{figure}
		\centering
		\includegraphics[width=10.5 cm]{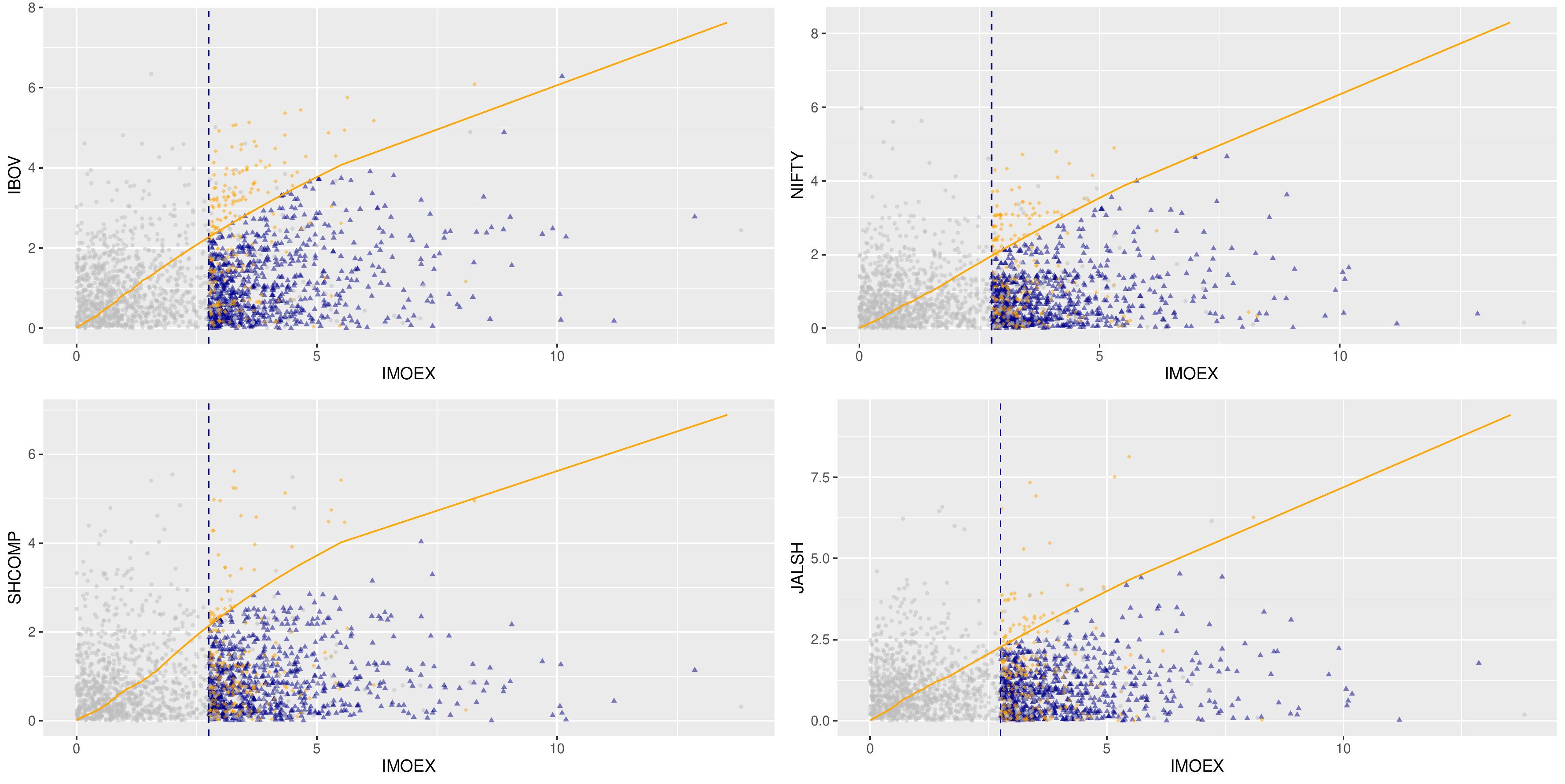}
		\caption[Prediction plot conditioning on IMOEX variable]{Prediction plot conditioning on IMOEX being above its $90^{th}$ percentile.}
		\label{bIBOVprdtng2}
	\end{figure}

\begin{figure}[H]
		\centering
		\includegraphics[width=10.5 cm]{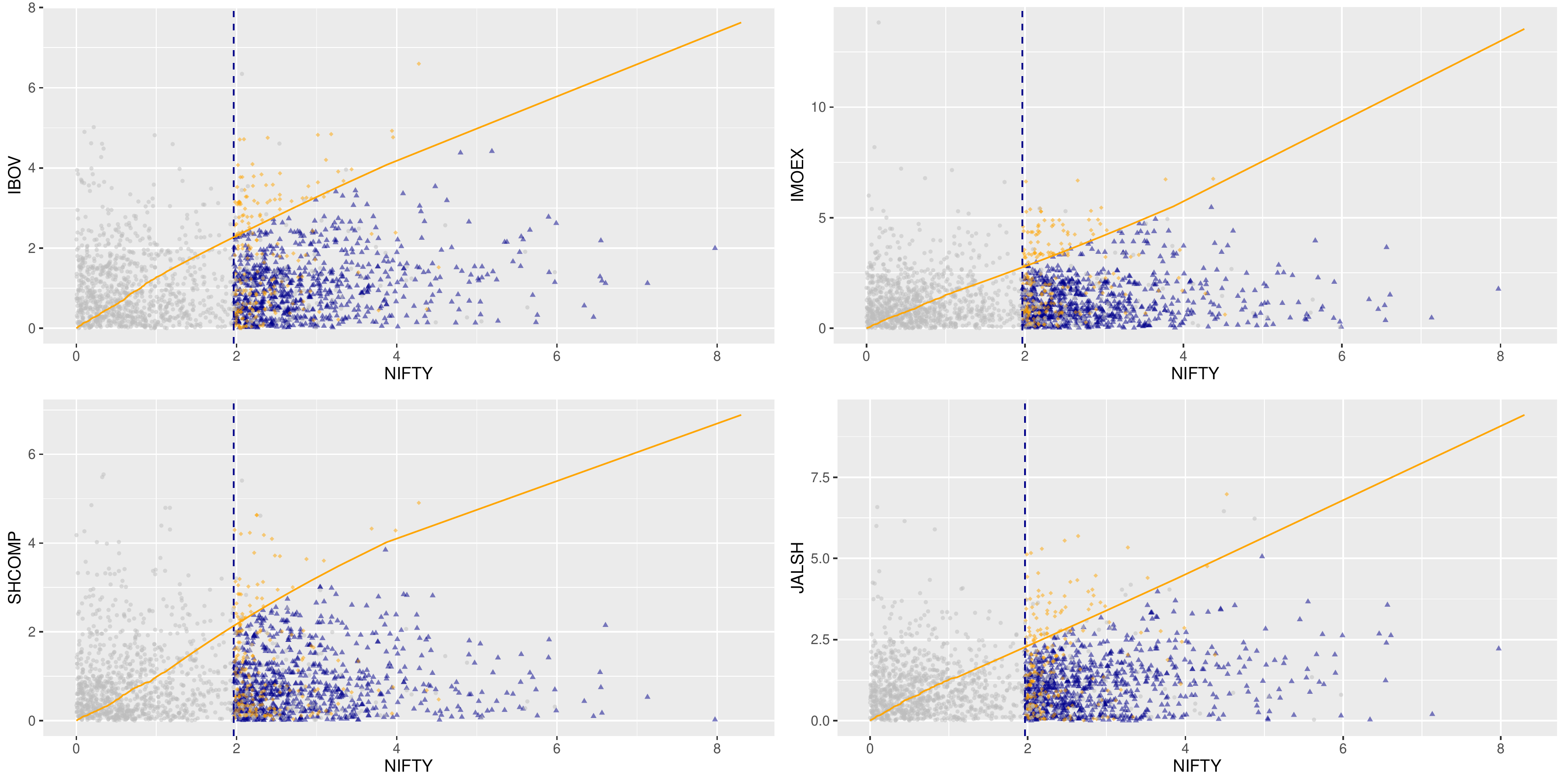}
		\caption[Prediction plot conditioning on NIFTY variable]{Prediction plot conditioning on NIFTY being above its $90^{th}$ percentile.}
		\label{bIBOVprdtng3}
		%\end{figure}
		\begin{flushleft}
		\end{flushleft}
		%\begin{figure}
		\centering
		\includegraphics[width=10.5 cm]{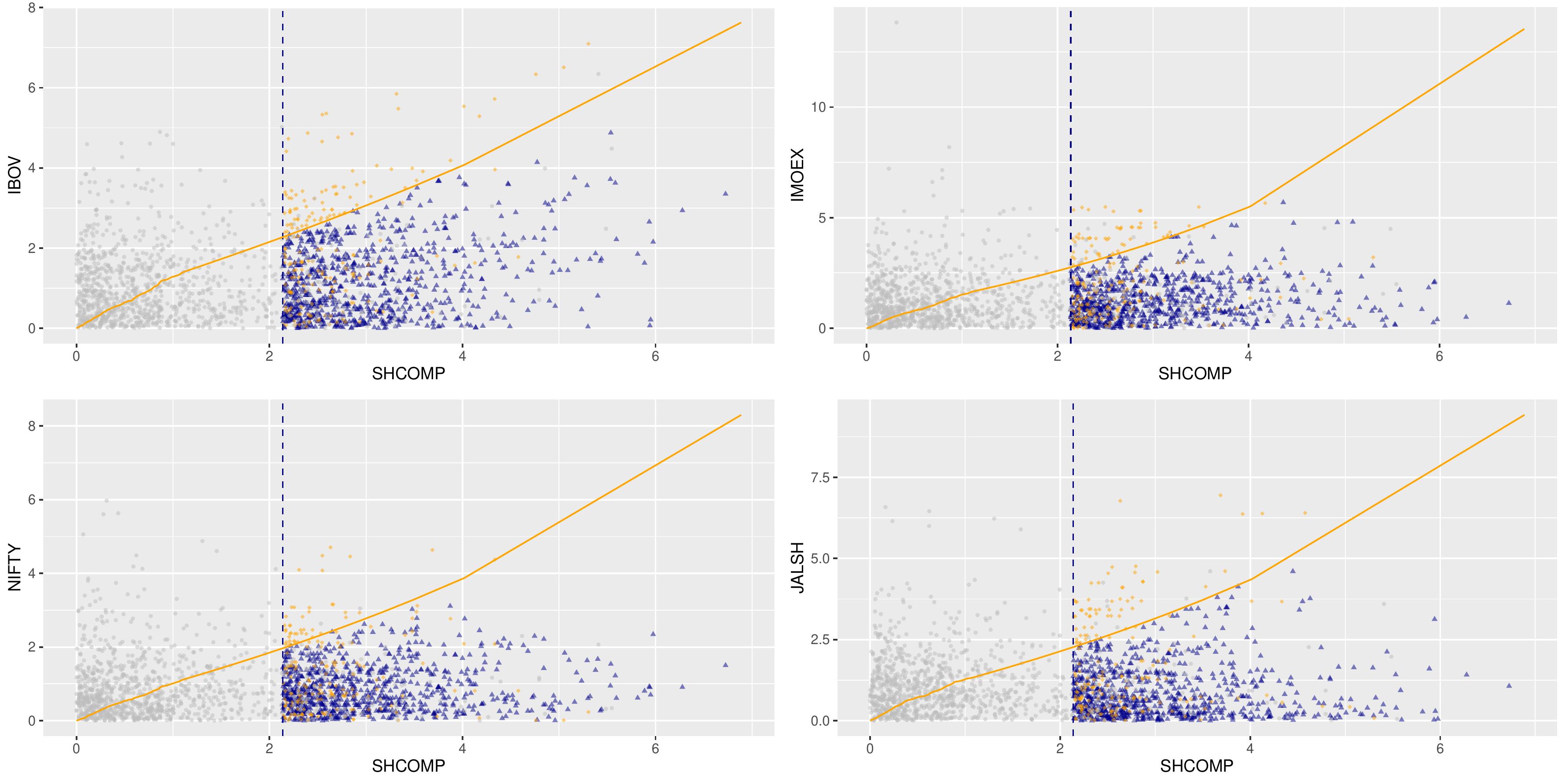}
		\caption[Prediction plot conditioning on SHCOMP variable]{Prediction plot conditioning on SHCOMP being above its $90^{th}$ percentile.}
		\label{bIBOVprdtng4}
	\end{figure}

\begin{figure}[H]
		\centering
		\includegraphics[width=10.5 cm]{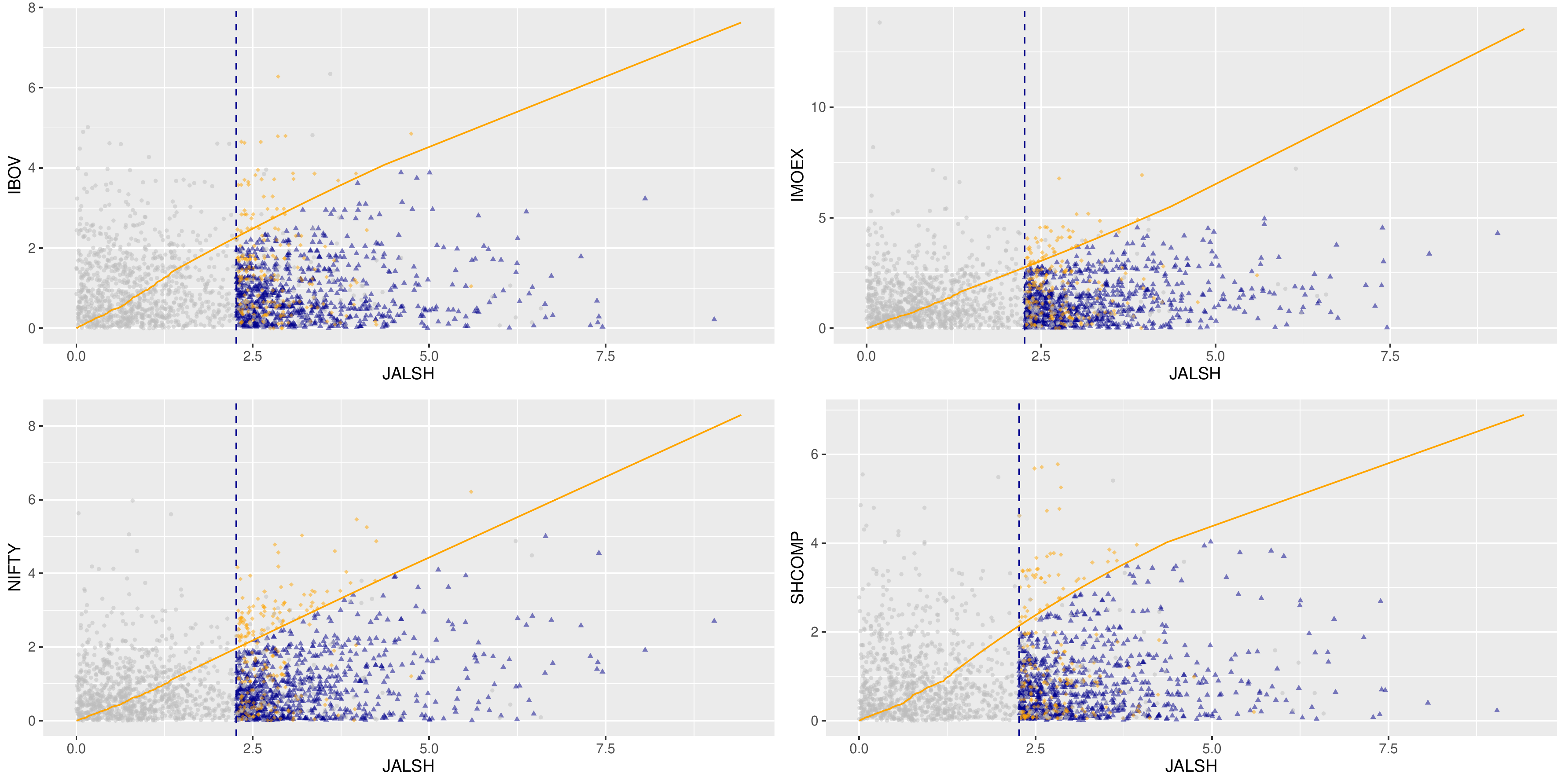}
		\caption[Prediction plot conditioning on JALSH variable]{Prediction plot conditioning on JALSH being above its $90^{th}$ percentile.}
		\label{bIBOVprdtng5}
	\end{figure}

	The prediction plots in Figures \ref{bIBOVprdtng} to \ref{bIBOVprdtng5} are
	used for a visual display of the CMEV model fit.
	
	\begin{table}[H]
		\centering
		\caption{Predicted conditional probability of threshold exceedance.}\label{esK1nmxc}
		\begin{tabular}{|l|l|l|l|l|}
			\hline
			Conditioning on    &   &  &  &\\
			IBOV    &  IMOEX & NIFTY & SHCOMP & JALSH\\                           \hline
			&  0.510 & 0.305 & 0.454 & 0.302\\
			&   &  & & \\                                     \hline
			Conditioning on  &   &  & & \\
			IMOEX   &  IBOV  & NIFTY & SHCOMP   & JALSH\\                          \hline
			&  0.433 & 0.346 & 0.358 & 0.351\\
			&  &  & & \\ \hline
			Conditioning on  &   &  & & \\
			NIFTY  &  IBOV  & IMOEX & SHCOMP & JALSH\\         \hline
			&  0.323 & 0.363 & 0.315 & 0.402\\
			&   &  &  &\\                         \hline
			Conditioning on   &  &  & &\\
			SHCOMP & IBOV  & IMOEX & NIFTY & JALSH \\                              \hline
			& 0.425 & 0.347 & 0.342 & 0.278\\
			&  &  &  &\\                                     \hline
			Conditioning on   &  &  & &\\
			JALSH &  IBOV  & IMOEX & NIFTY & SHCOMP \\                               \hline
			& 0.280 & 0.337 & 0.408 & 0.318\\
			\hline
		\end{tabular}
	\end{table}

	The predictions under the CMEV model fit are made by
	importance sampling through simulating the values of the other four
	(remaining) market-variables, given the conditioning variable being above a
	large prediction quantile (see \cite{southworth2016}). To choose a
	suitable prediction quantile, different candidate quantiles of $80^{th}, 90^{th},
	95^{th}$ and $99^{th}$ percentiles were examined, and the 90th percentile was found to be
	the most appropriate considering bias and variance trade-off. That is, the
	highest 10\% of the conditioning variable's values are being used. It should
	be noted that importance samples are known to have few values in the
	conditional tails (\cite{southworth2020}), hence the higher the prediction
	quantiles, the fewer the values in the conditional tails. This reason also
	makes the $90^{th}$ percentile a more suitable choice in preference to the higher
	variance $95^{th}$ and $99^{th}$ percentiles. As for the other four market-variables
	(excluding the conditioning variable), any value of threshold quantile can be
	used for the prediction (\cite{southworth2016}), and we used 0.7-quantiles
	or $70^{th}$ percentiles since the diagnostic plots of the lowess curves are
	smoothest for the dependence modelling at these quantiles.
	
	Having obtained the required prediction quantile, values of the
	market variables are simulated conditional on each of the five variables (in
	turn) being above its $90^{th}$ percentile. For instance, with the Brazilian IBOV
	market, values of the Russian IMOEX, Indian NIFTY, Chinese SHCOMP and South
	African JALSH market variables are simulated conditional on the Brazilian
	IBOV variable being above its $90^{th}$ percentile. Table \ref{esK1nmxc} shows the
	outcomes of the conditional distributions, i.e., the predicted (estimated)
	quantiles or probabilities of threshold exceedances.
	
	The prediction plots given in Figures \ref{bIBOVprdtng} to \ref{bIBOVprdtng5} are
	used for a visual display of the CMEV model fit and its prediction using
	importance samples. The plots show grey circles denoting the original
	data and data importance samples, represented by the blue triangles and
	orange diamonds, under the fitted CMEV model above the threshold for
	forecasting. The orange solid curves or lines in the plots are for references
	and each curve joins equal quantiles of the BRICS marginal distributions. The
	markets' paired variables with a perfect dependence will lie precisely on
	this line, comparable to a QQ plot's diagonal line.
	However, the curve is not a straight line because the two paired margins are
	not equal. From Figure \ref{bIBOVprdtng} for instance (by conditioning on the
	Brazilian IBOV market), it can be observed, when evaluated on a common
	quantile scale that all the blue triangles and orange diamonds are large in
	the (conditioning) IBOV variable, but only those displayed by the blue
	triangles are the largest in the IBOV variable, while the orange diamonds are the largest
	in each of the remaining four market variables. This visual scenario of the
	conditioning IBOV market-variable in Figure \ref{bIBOVprdtng} is also
	experienced by the remaining four conditioning markets, i.e., the IMOEX,
	NIFTY, SHCOMP and JALSH market-variables in Figures \ref{bIBOVprdtng2},
	\ref{bIBOVprdtng3}, \ref{bIBOVprdtng4} and \ref{bIBOVprdtng5}, respectively.
	
	Now for the prediction, we begin with conditioning on the Brazilian IBOV
	market in Figure \ref{bIBOVprdtng} and observed that contrary to the outcomes
	of the dependence structure in Table \ref{LondyMyWife} where the extremal
	dependence between market pair Brazilian IBOV and Russian IMOEX is the
	strongest, followed by the dependence between markets Brazilian IBOV and
	Chinese SHCOMP, the future relationship as shown in the figure, is predicted
	to be approximately strongest between market pair Brazilian IBOV and Indian
	NIFTY, followed by the pair of Brazilian IBOV and South African JALSH
	markets. This predicted extremal dependence within each pair of these latter
	markets (i.e., the Brazilian IBOV and Indian NIFTY, and Brazilian IBOV and
	South African JALSH) is better than it is in the two market-pairs Brazilian
	IBOV and Russian IMOEX, and Brazilian IBOV and Chinese SHCOMP in terms of
	near-linearity into the future, as displayed in the figure.
	
	Next, by conditioning on the Russian IMOEX market in Figure
	\ref{bIBOVprdtng2}, it is also observed that as opposed to the results of the
	dependence structure in Table \ref{LondyMyWife}, where the extremal
	dependence between market pair Russian IMOEX and Brazilian IBOV is the
	strongest, followed by the dependence between markets Russian IMOEX and
	Chinese SHCOMP, the future predicted relationship is strongest between
	Russian IMOEX and South African JALSH markets, followed by the dependence
	between pair Russian IMOEX and Indian NIFTY markets as displayed in the
	figure due to the sampled data that follow closely the curves of equal
	marginal quantiles.
	
	By conditioning on the Indian NIFTY market in Figure \ref{bIBOVprdtng3}, the
	evidence of strongest extremal dependence between market pair Indian NIFTY
	and South African JALSH, followed by the pair of Indian NIFTY and Brazilian
	IBOV markets, as shown in Table \ref{LondyMyWife} are extended into
	future dependence predictions as shown in the figure. That is, from the
	figure, the market pair Indian NIFTY and South African JALSH has the
	strongest predicted strength of direct proportionality (followed by the
	Indian NIFTY and Brazilian IBOV pairwise combination) than the rest of the
	paired markets.
	
	Next, by conditioning on the Chinese SHCOMP market in Figure
	\ref{bIBOVprdtng4}, it is also observed that the strongest extremal
	dependence between Chinese SHCOMP and Brazilian IBOV markets as shown by the
	result of the dependence structure in Table \ref{LondyMyWife} is carried into
	future extremal dependence as demonstrated by the near-linearity of their diagonal
	line in the prediction plot. Lastly, when conditioning on the South African
	JALSH market in Figure \ref{bIBOVprdtng5}, the pattern of the strongest
	extremal dependence between the pair of South African JALSH and Indian NIFTY
	markets, followed by the market pair South African JALSH and Russian IMOEX as
	shown in Table \ref{LondyMyWife} is further projected into the future as
	displayed in the figure. That is, predictions based on the levels of future
	relationship (extremal dependence) are greatest in these pairs of markets,
	with the latter pair following the former.
	
	\subsection{Bivariate point process modelling}
	The bivariate point process is now used for the extremal dependence modelling
	via the six parametric models that include the logistic, bilogistic,
	Husler-Reiss, negative logistic, negative bilogistic and the Dirichlet (or
	Coles-Tawn abbreviated ``ct") dependence models. The model that gives the best fit for the dependence
	structure in each of the ten paired markets is selected. Decisions relating to
	model selection (or comparison) for dependence structure of nested models can
	be addressed using standard likelihood ratio tests (\cite{coles1991}) and analysis of variance (ANOVA) (\cite{stephenson2018}), whereas
	for non-nested models, analytic goodness of fit statistic like the Akaike
	information criterion (AIC) can be used (\cite{coles1991,stephenson2018}). Hence, the AIC as used in the applied package ``evd" (\cite{stephenson2018}) is used for the model selection in each pair of the markets.
	
	Tables \ref{PPdepcen1vbn}, \ref{PPdepcen1} and \ref{PPdepecn2} show the
	dependence parameters $\alpha$ and $\beta$ along with their standard error
	likelihood estimates in parentheses. The logistic, negative logistic and
	Husler-Reiss models have a single dependence parameter $\alpha$. Therefore
	the $\beta$ is tabulated as ``Nil". The model with the lowest
	AIC value denotes the best fitting model for the dependence estimate of each
	pair of the BRICS markets.
	
	\begin{table}[H]
		\centering
		\caption{Estimates of the point process dependence modelling.}\label{PPdepcen1vbn}
		\begin{tabular}{|l|c|c|c|}
			\hline
			$\mathbf{Brazilian~IBOV~and}$ &  &  &\\
			$\mathbf{Russian~IMOEX}$    &  &  &  \\                     \hline
			$\mathbf{Parametric~model}$ & $\alpha$ & $\beta$ & AIC \\   \hline
			Logistic & 0.5994 (0.0117) & Nil & 3762.38 \\                       \hline
			Negative & 0.9055 (0.0314) & Nil & 3711.77 \\
			logistic &  &  &  \\                     \hline
			Husler-Reiss& 1.3413 (0.0345) & Nil & 3669.11 \\                     \hline
			Bilogistic & 0.5829 (0.0258) & 0.6155 (0.0247) & 3763.84 \\                   \hline
			Negative & 1.1938 (0.1134) & 1.0210 (0.0978) & 3712.99 \\
			bilogistic &  &  &  \\                  \hline
			%  Dirichlet & 0.8195 (0.1010) & 1.0172 (0.1369) & 3718.47 \\                      \hline
			ct (or Dirichlet) & 0.8195 (0.1010) & 1.0172 (0.1369) & 3718.47 \\                      \hline
			%   &  &  &  \\
			$\mathbf{Brazilian~IBOV~and}$ &  &  &\\
			$\mathbf{Indian~NIFTY}$      &  &  &  \\                               \hline
			$\mathbf{Parametric~model}$ & $\alpha$ & $\beta$ & AIC \\   \hline
			Logistic & 0.6066 (0.0114) & Nil & 3756.67 \\                       \hline
			Negative & 0.8854 (0.0300) & Nil & 3701.53 \\
			logistic &  &  &  \\                     \hline
			Husler-Reiss& 1.3174 (0.0330) & Nil & 3656.31 \\                     \hline
			Bilogistic & 0.6209 (0.0244) & 0.5909 (0.0267) & 3758.23 \\                   \hline
			Negative & 1.0470 (0.1029) & 1.2140 (0.1141) & 3702.85 \\
			bilogistic &  &  &  \\                  \hline
			ct (or Dirichlet) & 0.9829 (0.1364) & 0.7940 (0.0959) & 3710.40 \\                      \hline
			%     &  &  &  \\
			$\mathbf{Brazilian~IBOV~and}$ &  &  &\\
			$\mathbf{Chinese~SHCOMP}$       &  &  &  \\                     \hline
			$\mathbf{Parametric~model}$ & $\alpha$ & $\beta$ & AIC \\   \hline
			Logistic & 0.6086 (0.0114) & Nil & 3788.63 \\                       \hline
			Negative & 0.8779 (0.0296) & Nil &  3733.10\\
			logistic &  &  &  \\                     \hline
			Husler-Reiss& 1.3100 (0.0326) & Nil & 3683.98 \\                     \hline
			Bilogistic & 0.6264 (0.0235) & 0.5901 (0.0250) & 3789.91 \\                   \hline
			Negative & 1.0629 (0.0988) & 1.2186 (0.1111) & 3734.46 \\
			bilogistic &  &  &  \\                  \hline
			ct (or Dirichlet) & 0.9515 (0.1229) & 0.7931 (0.0949) & 3742.99 \\                      \hline
			%   &  &  &  \\
			$\mathbf{Brazilian~IBOV~and}$ &  &  &\\
			$\mathbf{S/African~JALSH}$       &  &  &  \\                     \hline
			$\mathbf{Parametric~model}$ & $\alpha$ & $\beta$ & AIC \\   \hline
			Logistic & 0.6132 (0.0114) & Nil & 3817.37 \\                       \hline
			Negative & 0.8646 (0.0292) & Nil &  3761.19\\
			logistic &  &  &  \\                     \hline
			Husler-Reiss& 1.2906 (0.0321) & Nil & 3715.79 \\                     \hline
			Bilogistic & 0.6137 (0.0246) & 0.6125 (0.0254) & 3819.37 \\                   \hline
			Negative & 1.1414 (0.1098) & 1.1707 (0.1096) & 3763.17 \\
			bilogistic &  &  &  \\                  \hline
			ct (or Dirichlet) & 0.8591 (0.1121) & 0.8303 (0.1027) & 3772.86 \\
			\hline
		\end{tabular}
	\end{table}
	
	\begin{table}[H]
		\centering
		\caption{Estimates of the point process dependence modelling.}\label{PPdepcen1}
		\begin{tabular}{|l|c|c|c|}
			\hline
			$\mathbf{Russian~IMOEX~and}$ &  &  &\\
			$\mathbf{Indian~NIFTY}$     &  &  &  \\                     \hline
			$\mathbf{Parametric~model}$ & $\alpha$ & $\beta$ & AIC \\   \hline
			Logistic & 0.6073 (0.0116) & Nil & 3873.50 \\                       \hline
			Negative & 0.8818 (0.0302) & Nil & 3821.21 \\
			logistic &  &  &  \\                     \hline
			Husler-Reiss& 1.3110 (0.0332) & Nil & 3780.47 \\                     \hline
			Bilogistic & 0.6185 (0.0240) & 0.5955 (0.0254) & 3875.22 \\                   \hline
			Negative & 1.0850 (0.1033) & 1.1839 (0.1100) & 3822.96 \\
			bilogistic &  &  &  \\                  \hline
			ct (or Dirichlet) & 0.9256 (0.1208) & 0.8203 (0.0997) & 3830.56 \\                      \hline
			%  &  &  &  \\
			$\mathbf{Russian~IMOEX~and}$ &  &  &\\
			$\mathbf{Chinese~SHCOMP}$     &  &  &  \\                               \hline
			$\mathbf{Parametric~model}$ & $\alpha$ & $\beta$ & AIC \\   \hline
			Logistic & 0.6072 (0.0116) & Nil & 3941.75 \\                       \hline
			Negative & 0.8806 (0.0303) & Nil & 3887.51 \\
			logistic &  &  &  \\                     \hline
			Husler-Reiss& 1.3145 (0.0334) & Nil & 3838.94 \\                     \hline
			Bilogistic & 0.6140 (0.0244) & 0.6003 (0.0252) & 3943.65 \\                   \hline
			Negative & 1.1083 (0.1068) & 1.1628 (0.1106) & 3889.43 \\
			bilogistic &  &  &  \\                  \hline
			ct (or Dirichlet) & 0.9066 (0.1206) & 0.8415 (0.1072) & 3897.57 \\                      \hline
			%    &  &  &  \\
			$\mathbf{Russian~IMOEX~and}$ &  &  &\\
			$\mathbf{S/African~JALSH}$         &  &  &  \\                     \hline
			$\mathbf{Parametric~model}$ & $\alpha$ & $\beta$ & AIC \\   \hline
			Logistic & 0.6126 (0.0114) & Nil & 3935.83 \\                       \hline
			Negative & 0.8653 (0.0293) & Nil &  3883.64\\
			logistic &  &  &  \\                     \hline
			Husler-Reiss& 1.2908 (0.0321) & Nil & 3839.97 \\                     \hline
			Bilogistic & 0.6068 (0.0232) & 0.6184 (0.0231) & 3937.75 \\                   \hline
			Negative & 1.1356 (0.1036) & 1.1759 (0.1065) & 3885.60\\
			bilogistic &  &  &  \\                  \hline
			ct (or Dirichlet) & 0.8727 (0.1101) & 0.8151 (0.0967) & 3894.70 \\                      \hline
			%   &  &  &  \\
			$\mathbf{Indian~NIFTY~and}$ &  &  &\\
			$\mathbf{Chinese~SHCOMP}$        &  &  &  \\                     \hline
			$\mathbf{Parametric~model}$ & $\alpha$ & $\beta$ & AIC \\   \hline
			Logistic & 0.6063 (0.0117) & Nil & 3805.05 \\                       \hline
			Negative & 0.8826 (0.0305) & Nil &  3758.71\\
			logistic &  &  &  \\                     \hline
			Husler-Reiss& 1.3070 (0.0332) & Nil & 3719.32 \\                     \hline
			Bilogistic & 0.6142 (0.0231) & 0.5980 (0.0245) & 3806.90 \\                   \hline
			Negative & 1.1027 (0.1029) & 1.1632 (0.1055) & 3760.61 \\
			bilogistic &  &  &  \\                  \hline
			ct (or Dirichlet) & 0.8996 (0.1146) & 0.8365 (0.1005) & 3767.56 \\
			\hline
		\end{tabular}
	\end{table}
	
	There are ten pairwise combinations of the five BRICS markets as presented in
	Tables \ref{PPdepcen1vbn}, \ref{PPdepcen1} and \ref{PPdepecn2}. The results
	from the tables show that the model that best describes all the ten paired
	markets is the Husler-Reiss, with the lowest AIC value in each pair. The Husler-Reiss model produces
	complete dependence between variables as $\alpha$ tends to $\infty$, while
	independence is obtained as $\alpha$ approaches 0. The tables show that
	maximization of likelihood under the Husler-Reiss model for the ten paired
	markets yields estimates of $1.2883 \leq \hat{\alpha} \leq 1.3413$, and
	corresponding standard errors of between 0.0320 and 0.0345 inclusive.
	
	\begin{table}[H]
		\centering
		\caption{Estimates of the point process dependence modelling.}\label{PPdepecn2}
		%\begin{tabular}{|l|l|l|l|}
		\begin{tabular}{|l|c|c|c|}
			\hline
			$\mathbf{Indian~NIFTY~and}$ &  &  &\\
			$\mathbf{S/African~JALSH}$         &  &  &  \\                     \hline
			$\mathbf{Parametric~model}$ & $\alpha$ & $\beta$ & AIC \\   \hline
			Logistic & 0.6040 (0.0115) & Nil & 3745.97 \\                       \hline
			Negative & 0.8924 (0.0304) & Nil &  3699.66\\
			logistic &  &  &  \\                     \hline
			Husler-Reiss & 1.3200 (0.0332) & Nil & 3664.18 \\                     \hline
			Bilogistic & 0.6054 (0.0229) & 0.6026 (0.0229) & 3747.97 \\                   \hline
			Negative & 1.1070 (0.0979) & 1.1343 (0.1005) & 3701.64 \\
			bilogistic &  &  &  \\                  \hline
			% Dirichlet & 0.8946 (0.1064) & 0.8649 (0.1018) & 3707.80 \\                      \hline
			ct (or Dirichlet) & 0.8946 (0.1064) & 0.8649 (0.1018) & 3707.80 \\                      \hline
			%  &  &  &  \\
			$\mathbf{Chinese~SHCOMP~and}$ &  &  &\\
			$\mathbf{S/African~JALSH}$             &  &  &  \\                     \hline
			$\mathbf{Parametric~model}$ & $\alpha$ & $\beta$ & AIC \\   \hline
			Logistic & 0.6149 (0.0114) & Nil & 3869.14 \\                       \hline
			Negative & 0.8605 (0.0290) & Nil &  3810.94\\
			logistic &  &  &  \\                     \hline
			Husler-Reiss & 1.2883 (0.0320) & Nil & 3764.18 \\                     \hline
			Bilogistic & 0.6283 (0.0242) & 0.6006 (0.0260) & 3870.76 \\                   \hline
			Negative & 1.1129 (0.1076) & 1.2121 (0.1143) & 3812.71 \\
			bilogistic &  &  &  \\                  \hline
			%   Dirichlet & 0.8926 (0.1183) & 0.7915 (0.0977) & 3823.01 \\
			ct (or Dirichlet) & 0.8926 (0.1183) & 0.7915 (0.0977) & 3823.01 \\
			\hline
		\end{tabular}
	\end{table}
	
	These dependence estimates $\hat{\alpha}$ tend to move further away from 0,
	and are therefore significantly different from independence but reasonably
	correspond to weak levels of dependence. However, since the larger (or
	smaller) the likelihood estimate of $\alpha$, the stronger (or weaker) the
	strength of dependence (see \cite{coles2001}), the dependence values of 1.3413 and
	1.3200 for the pairs of Brazilian IBOV and Russian IMOEX, and Indian NIFTY
	and South African JALSH markets, respectively, as shown in Tables
	\ref{PPdepcen1vbn} and \ref{PPdepecn2} can be classified as approximately
	fairly strong. These findings are consistent with the results obtained
	from the CMEV modelling. The only likely exception to the consistency is
	between the pair of Brazilian IBOV and Chinese SHCOMP markets, which has a
	fairly strong dependence under the CMEV modelling, but produced a nearly weak
	dependence under the point process.
	
	\begin{figure}[H]
		\centering
		\includegraphics[height=4.0in, width=5.0in]{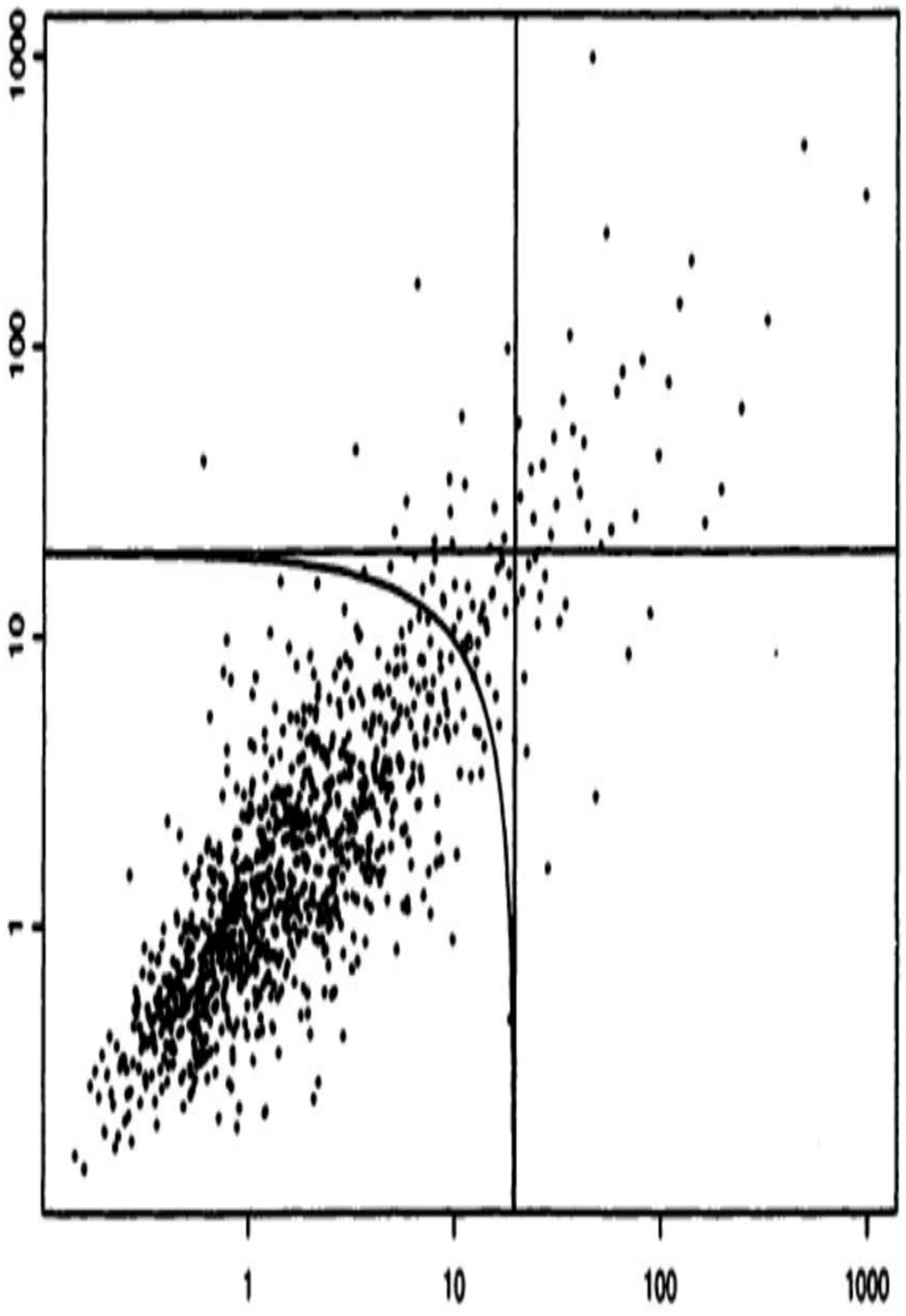}
		\caption[The CMEV and bivariate point process models' thresholds]{The CMEV and bivariate point process models' same points thresholds. Source \cite{coles2001}.}
		\label{MAMAxn4wevazm}
	\end{figure}
	
	\subsubsection{Point process and CMEV models compared}
	This section compares the outcomes and modelling approaches of the bivariate
	point process and the CMEV models. An adequately objective comparison of the
	dependence approach of the point process with the threshold excess method of
	the CMEV model is made by using the dependence threshold $u$ that corresponds
	to the 0.7-quantile used for the CMEV modelling. This is made
	possible when a threshold $u$ is selected where its intersection with the
	axes takes place at the same points for both models, as exemplified in Figure
	\ref{MAMAxn4wevazm} (see \cite{coles2001}). At these points (in this study), the
	thresholds for the point process and threshold excess approach of the CMEV
	models are selected to intersect the axes at the marginal 0.7-quantiles.
	
	\begin{table}[H]
		\centering
		\caption{Extremal dependence structures of the CMEV and point process models.}\label{PcomJERUecn2}
		\begin{tabular}{|l|c|c|}
			\hline
			&  &  \\
			$\textbf{Panel A}$ &  &  \\ \hline
			$\textbf{Paired markets}$ & $\textbf{CMEV model}$ & $\textbf{Bivariate point process model}$ \\ \hline
			Brazilian IBOV and &  &  \\
			Indian NIFTY & Weak & Weak \\ \hline
			Brazilian IBOV and &  &  \\
			S/African JALSH & Weak & Weak \\ \hline
			Russian IMOEX and &  &  \\
			Indian NIFTY & Weak & Weak \\ \hline
			Russian IMOEX and &  &  \\
			Chinese SHCOMP & Weak & Weak \\ \hline
			Russian IMOEX and &  &  \\
			S/African JALSH & Weak & Weak \\ \hline
			Indian NIFTY and &  &  \\
			Chinese SHCOMP & Weak & Weak \\ \hline
			Chinese SHCOMP and &  &  \\
			S/African JALSH & Weak & Weak \\ \hline
			&  &  \\
			\textbf{Panel B} &  &  \\ \hline
			Brazilian IBOV and &  &  \\
			Russian IMOEX & Fairly strong & Fairly strong \\ \hline
			Indian NIFTY and &  &  \\
			S/African JALSH & Fairly strong & Fairly strong \\ \hline
			&  &  \\
			\textbf{Panel C} &  &  \\ \hline
			Brazilian IBOV and &  &  \\
			Chinese SHCOMP & Fairly strong & Nearly weak \\
			%  &  &  \\
			\hline
		\end{tabular}
	\end{table}
	
	The difference between the two models is due to the different regions where
	their approximations assume validity. From Figure \ref{MAMAxn4wevazm}, the
	limit of the point process model is assumed to be valid in the entire region
	above the curve, to the right of the boundary, i.e., the boundary of the
	threshold and the axes. The point process can use exceedance
	observations above the curve in the two margins and the joint
	upper (right) quadrant. This upward region of the boundary or curve is
	considered appropriately extreme for the limit outcome of the point process
	to offer a valid approximation. As opposed to this, given the same marginal
	threshold as the point process, the limit of the threshold excess method of
	the CMEV model takes accuracy only in the joint upper (right) quadrant
	(\cite{coles2001}).
	
	Hence, since every extreme observation above the curved region is used, the
	point process can model many more points or exceedances that
	contribute to the likelihood estimation. It gives more information than
	the threshold excess method of the CMEV model. Table \ref{PcomJERUecn2}
	summarily presents the extremal dependence structures of the ten paired
	markets under the two models: the CMEV and bivariate point process models. It
	can be observed, as shown in panel A of the table, that asymptotic dependence
	is weak but significant for seven out of the ten pairs of the markets under
	both models. Next, panel B displays a fairly strong dependence outcome for
	two paired markets under the two comparative models. In panel C, however, the
	two models, as discussed earlier, yielded a slightly different outcome, with
	the CMEV model resulting in a fairly strong dependence. At the same time, the point
	process produced a nearly weak dependence for the same paired markets. Hence,
	there is consistency in the modelling capabilities of the two models for nine
	out of the ten paired markets and near consistency in the tenth market pair.
	
	\section{Discussion}
	One barrier to creating wealth is the risk of extreme
	losses, and any portfolio that seeks wealth maximisation should mitigate this
	risk. Risk is defined as the instability or unpredictability of
	returns from an investment. Intensely advanced risk management systems and
	rigorous research have been conducted to comprehend and tackle
	this risk. The most robust outcome from the works is rather
	straightforward: diversify.
	
	Creating a well-diversified portfolio within an asset class and across
	classes of assets, and also geographically by investing domestically and in
	foreign markets can help investors to deal with returns variability, thereby
	forestalling extreme losses. Diversification is a  risk management approach
	where a wide variety of investments are mixed within a portfolio. Although
	portfolio diversification may limit investment gains in the short-term, but
	in the long-term, it decreases portfolio risk, hedges against volatility in
	markets, and gives higher returns (\cite{segal2021}).
	
	Diversification is fundamental for creating portfolios that maximise
	return for a certain risk level or, on the contrary, minimise risk for a
	certain return level. Reisen \cite{reisen2000} in his findings revealed that risk is
	easier to reduce via international diversification than through domestic
	diversification. Chen \cite{chen2020} further showed that having a global
	portfolio can be used for risk reduction in investment. This is because if
	equities under perform in one nation’s domestic market, a gain in the
	international holdings for the investor in any other country (s) can smooth out
	the return. Hence, through global investment, an investor can
	raise returns through risk reduction.
	
	Amongst others, Solnik \cite{solnik1974} indicated that international (portfolio)
	diversification enables investors to achieve a greater efficient edge when
	compared to diversifying domestically (\cite{aloui2011}). Moreover, because
	of different structures of industry in other nations and since different
	economies do not precisely adhere to a similar business cycle, diversifying
	across countries whose economic (or trade) cycles do not have perfect
	correlation can typically reduce returns variability for investors and
	portfolio managers (\cite{solnik1974,zonouzi2014}). This is so because
	different factors drive each nation's market at any given time. More
	directly, the benefits of diversification are obtained from risk reduction in
	nearly uncorrelated and negatively correlated markets (\cite{odit2011}).
	This, in other words, would mean the existence of low correlations of stock
	returns between different economies (\cite{solnik1974}).
	
	Yavas \cite{yavas2007} addressed the subject of risk reduction via international
	diversification by revealing that diversification across nations within an
	industry can produce much more effective risk reduction than industry
	diversification within a country.  If an entire sector fails in
	one nation but succeeds in another, investing in the same industry in both
	nations may well hedge the risk. Numerous possible benefits like the
	mentioned risk reduction and increase in returns have led to investors
	internationalizing their portfolios. These apparent benefits are seen by
	\cite{bartram2001} as essential motivations needed for international
	portfolio investments.
	
	Results in Tables \ref{LondyMyWife} and \ref{PPdepcen1vbn} –
	\ref{PcomJERUecn2} describe the weak (or low) positive and negative extremal
	dependence associations and the fairly strong (but still low) relationship
	exhibited by the pairs of the BRICS markets. All these outcomes from the ten
	pairwise combinations of the BRICS stock markets signify varying levels of low
	extremal dependence.  With low correlation or simply weak asymptotic
	dependence and negative extremal dependence being the fundamental
	requirements for efficient portfolio diversification, investors and all
	market participants can seize the rich investment opportunities presented by
	these markets as presented in the tables. The weak extremal dependence
	indicates that extreme losses in one market do not easily spill over to the
	other markets. On the other hand, the negative asymptotic dependence (for
	instance, in the four paired markets under the CMEV modelling in Table
	\ref{LondyMyWife}) means that under-performance or negative outcomes caused by
	extreme losses in one market can be offset or hedged by good (positive)
	returns performance in the other market for a paired market.
	
	To begin with, the findings as shown in panel A of the summary Table
	\ref{PcomJERUecn2} show weak extremal (asymptotic) dependence between each of
	the seven (out of ten) paired markets. That is, from the table, beneficial
	risk reduction and high investment returns through an international portfolio
	diversifications can be derived from any of the following as indicated in
	panel A: 1) if a Brazilian investor invests in the domestic market and any of
	the Indian and South African markets; 2) if a Russian investor jointly
	invests in the domestic market and any of the Indian, Chinese, and South
	African market; 3) if an Indian investor invests in the domestic market and
	in the Chinese market; 4) if a Chinese investor invests in the domestic
	market and in the South African market. Moreover, an authorised international
	investor outside any of these countries can likewise invest comfortably and
	obtain good returns from any of the markets' combinations.
	
	Next, in panel B (that contains two paired markets), a fairly good investment
	opportunity derivable from international portfolio diversifications can also
	be expected. This is so because the extremal dependence between the markets
	in these market pairs is ``fairly strong" compared to the ``weak asymptotic."
	dependence for panel A markets. Hence the diversification benefits are meant
	to be lower for these two paired markets when compared to the seven paired
	markets in panel A. Following these outcomes in panel B, 1) a Brazilian
	investor can earn fairly good returns by diversifying in the domestic
	market and the Russian stock market, and 2) an Indian investor can also
	obtain beneficially adequate portfolio diversification and investment returns
	jointly from the domestic market and the South African stock market. This
	opportunity is also open to international investors interested in
	portfolio investments in these market pairs.
	
	The findings in panel C for the tenth paired markets are not outrightly
	direct since the two models (CMEV and bivariate point process) produced
	slightly different outcomes. However, from the table, these two outcomes
	(fairly strong and nearly weak extremal dependencies) are still categorised
	approximately under low asymptotic dependence for any investor with
	investment interest in these markets. Hence, from panel C, a Brazilian
	investor can derive a beneficial or fairly beneficial investment return
	the outcome of diversifying in the domestic market and the Chinese stock
	market.
	
	It can be observed from the findings in Tables \ref{LondyMyWife},
	\ref{PPdepcen1vbn} – \ref{PcomJERUecn2} that investors will possibly achieve
	the most desirable portfolio diversification benefits in the Chinese$-$South
	African combination with the lowest extremal dependence and the least
	desirable diversified portfolio investment in the Brazilian$-$Russian
	combination with the highest asymptotic dependence. The summary extremal
	dependence results in panels A, B and C of Table \ref{PcomJERUecn2} are
	highly beneficial to investors, traders, portfolio managers and other markets
	participants who are interested in maximising their investment returns and
	financial gains, and in the process, mitigate possible investment downturns
	through international portfolio diversification in the BRICS stock markets.
	
	It is also important to know that these days, unlike in the past,
	investors can effectively build internationally diversified portfolios
	through mutual funds and international exchange-traded funds (ETFs) which
	focus on handling foreign equities, with a quite reasonable and quick way to
	diversify (\cite{yavas2007,kuepper2019,chen2020}).
	
	Similar studies which have modelled the extremal dependence of the BRICS stock markets include those of \cite{mensi2016,ijumba2013,lee2017,afuecheta2020,babu2015}, among others. However, none of the authors has used the combined multivariate versions of
	the point process models through the logistic, negative logistic,
	Husler-Reiss, Bilogistic, negative bilogistic and Coles-Tawn (or Dirichlet)
	models, and the CMEV model before this study to the best of the authors'
	knowledge. Hence, this study robustly models and estimates the extremal dependencies in the ten pairs of the BRICS stock markets. The following statistical R packages were used in the study: "evd," "texmex" and "evmix."
	
	%%%%%%%%%%%%%%%%%%%%%%%%%%%%%%%%%%%%%%%%%%%
	\section{Conclusions}
	
	In this study, we used data from the five BRICS stock markets with emphasis
	on modelling the extremal dependence in the pairwise combinations
	of the markets using a multivariate approach.
	
	Although evidence from the literature suggests that some studies have been
	done on modelling volatility of the BRIC(S) markets (see \cite{lee2017}) and a few
	studies on modelling their interdependence or co-movement (see \cite{ijumba2013,babu2015}), no evidence is
	currently available on modelling their extremal dependence using the
	conditional extreme value (CEV) model and point process approach. This is the
	the gap this study bridged by modelling the risk in each market and the
	asymptotic dependence of the paired markets using the CEV model and point
	process approach. 
	
	The two EVT models of conditional extreme value (CEV) and point process are adequately
	satisfactory for modelling the risk, based on the results of various diagnostics and
	tests under the univariate modelling. For extremal dependence modelling, however,
	the bivariate point process was able to model many more extreme observations or exceedances
	that contribute to the likelihood estimation and it gives more information
	than the threshold excess method of the CMEV model.
	
	A limitation of this study is that we only used MLE for estimating the parameters of the developed models. Future research can use the Bayesian parameter estimation approach and then compare its results with that of the benchmark MLE. Another research direction would be to analyse the impact of the current war between Russia and Ukraine on the BRICS' equity markets.
	
	%This section is not mandatory, but can be added to the manuscript if the discussion is unusually long or complex.
	
	%%%%%%%%%%%%%%%%%%%%%%%%%%%%%%%%%%%%%%%%%%%
	%\section{Patents}
	%
	%This section is not mandatory, but may be added if there are patents resulting from the work reported in this manuscript.
	%
	%%%%%%%%%%%%%%%%%%%%%%%%%%%%%%%%%%%%%%%%%%%
	\vspace{6pt} 
	
	%%%%%%%%%%%%%%%%%%%%%%%%%%%%%%%%%%%%%%%%%%
	%% optional
	%\supplementary{The following are available online at \linksupplementary{s1}, Figure S1: title, Table S1: title, Video S1: title.}
	
	
	%%%%%%%%%%%%%%%%%%%%%%%%%%%%%%%%%%%%%%%%%%
%	%%%%%%%%%%%%%%%%%%%%%%%%%%%%%%%%%%%%%%%%%%
%	\authorcontributions{The results of this study were obtained from a submitted doctor of philosophy thesis at the University of Venda.\\	Conceptualization, R.M. and C.S.; methodology, C.S. and R.M.; software, C.S. and R.M.; validation, C.S., R.M., W.C. and W.G.; formal analysis, C.S., R.M., investigation,  C.S., R.M., W.C. and W.G.; data curation, C.S. and R.M.; writing---original draft preparation, C.S. and R.M.; writing---review and editing,  C.S., R.M., W.C. and W.G.; supervision, C.S., R.M., W.C. and W.G.; project administration, C.S., W.C. and W.G. All authors have read and agreed to the published version of the manuscript.}
%	
%	\funding{Not applicable.}
%	
%	\institutionalreview{Not applicable.}
%	
%	\informedconsent{Not applicable.}
%	
%	\dataavailability{The data used in this study are from  , website (\url{https://},
%		accessed on 17 June 2020).} %MDPI:Please add accessed date. ADDED
%	
	\section*{Acknowledgments}
	The authors acknowledge the Southern African Universities Radiometric Network (SAURAN) for providing the data.	
	
%	\conflictsofinterest{The authors declare no conflict of interest.} 
	
	%%%%%%%%%%%%%%%%%%%%%%%%%%%%%%%%%%%%%%%%%%
	%% Optional
\section*{Abbreviations}
		The following abbreviations are used in this manuscript:\\
		
		\noindent 
		\begin{tabular}{@{}ll}
			BRICS & Brazil, Russia, India, China and South Africa \\
			GARCH & Generalised Autoregressive   Conditional Heteroskedasticity \\
			CMEV & Conditional Multivariate Extreme Value \\
			GAS & Generalised Autoregressive Score \\
			GFC & Global Financial Crisis\\
			EVT & Extreme Valu Theory \\
			GEVD & Generalised Extreme Value Distribution \\
			GPD & Generalised Pareto Distribution \\
			MLE & Maximum Likelihood Estimation \\
	\end{tabular}
	
	\section*{Appendices}

	\subsection*{Appendix A1: Box plots of the markets returns data}
	
	%Figure \ref{boxplotsreturns} shows the box plots of the stock market returns of the BRICS countries. The distributions for all the markets appear to be symmetrical with long tails. 
	\begin{figure}[H]
		\centering
		\includegraphics[width=14cm]{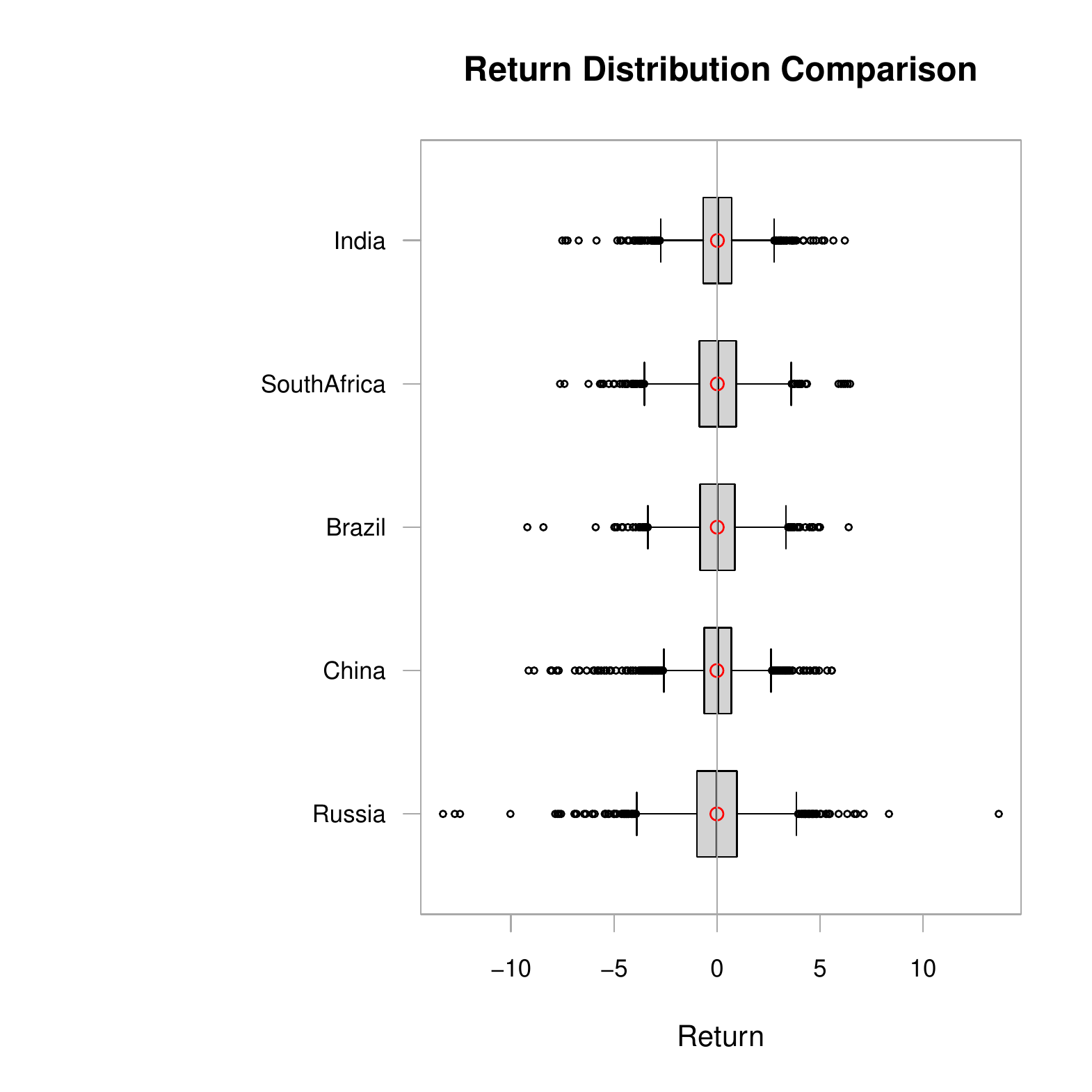}
		\caption[Box plots of the markets returns data]{Box plots of the markets returns data.}
		\label{boxplotsreturns}
	\end{figure}	
	
	%\subsection*{Appendix A2: Marginal model diagnostics}
	%
	%
	%\begin{figure}[H]
	%\centering
	%\includegraphics[width=10.5 cm]{figures/IBOV-AND-OTHERS-70TH-PERCENTILE.pdf}
	%\caption[Marginal model diagnostics for IBOV variable]{Marginal model diagnostics for IBOV variable.}
	%\label{margQUA}
	%%\end{figure}
	%\begin{flushleft}
	%\end{flushleft}
	%%\begin{figure}
	%\centering
	%\includegraphics[width=10.5 cm]{figures/IMOEX-AND-OTHERS-80TH-PERCENTILE.pdf}
	%\caption[Marginal model diagnostics for IMOEX variable]{Marginal model diagnostics for IMOEX variable.}
	%\label{margQUA1}
	%\end{figure}
	%
	%\begin{figure}[H]
	%\centering
	%\includegraphics[width=10.5 cm]{figures/NIFTY-AND-OTHERS-70TH-PERCENTILE.pdf}
	%\caption[Marginal model diagnostics for NIFTY variable]{Marginal model diagnostics for NIFTY variable.}
	%\label{margQUA2}
	%%\end{figure}
	%\begin{flushleft}
	%\end{flushleft}
	%%\begin{figure}
	%\centering
	%\includegraphics[width=10.5 cm]{figures/SHCOMP-AND-OTHERS-70TH-PERCENTILE.pdf}
	%\caption[Marginal model diagnostics for SHCOMP variable]{Marginal model diagnostics for SHCOMP variable.}
	%\label{margQUA3}
	%\end{figure}
	%
	%\begin{figure}[H]
	%\centering
	%\includegraphics[width=10.5 cm]{figures/JALSH-AND-OTHERS-70TH-PERCENTILE.pdf}
	%\caption[Marginal model diagnostics for JALSH variable]{Marginal model diagnostics for JALSH variable.}
	%\label{margQUA4}
	%\end{figure}
	%
	
	\subsection*{Appendix A2: Dependence model diagnostics}
	\begin{figure}[H]
		\centering
		\includegraphics[width=10.5 cm]{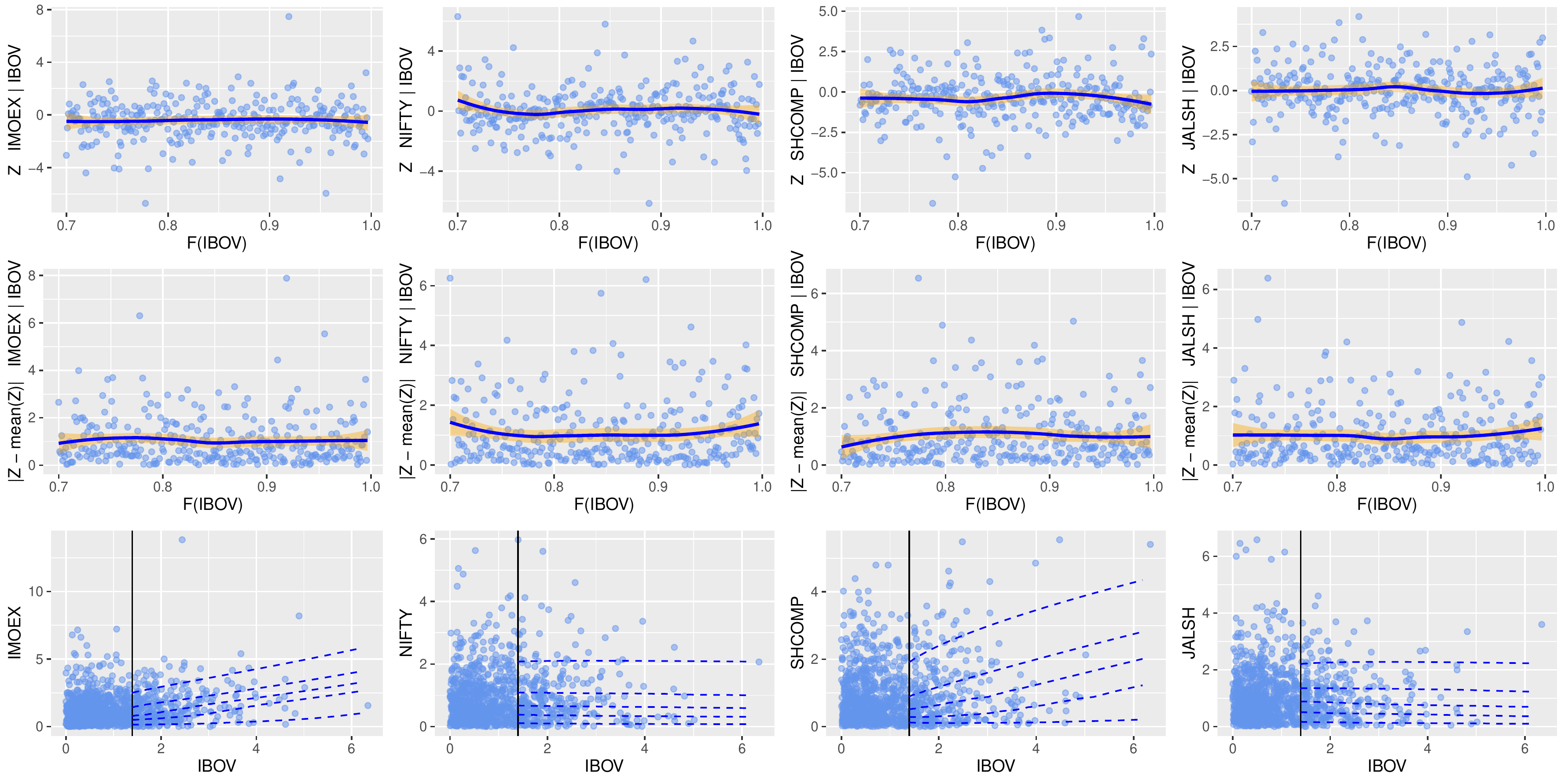}
		\caption[Dependence model diagnostics: conditioning on the IBOV variable]{Dependence model diagnostics: conditioning on the IBOV variable.}
		\label{DepQuaq1}
		%\end{figure}
		\begin{flushleft}
		\end{flushleft}
		%\begin{figure}
		\centering
		\includegraphics[width=10.5 cm]{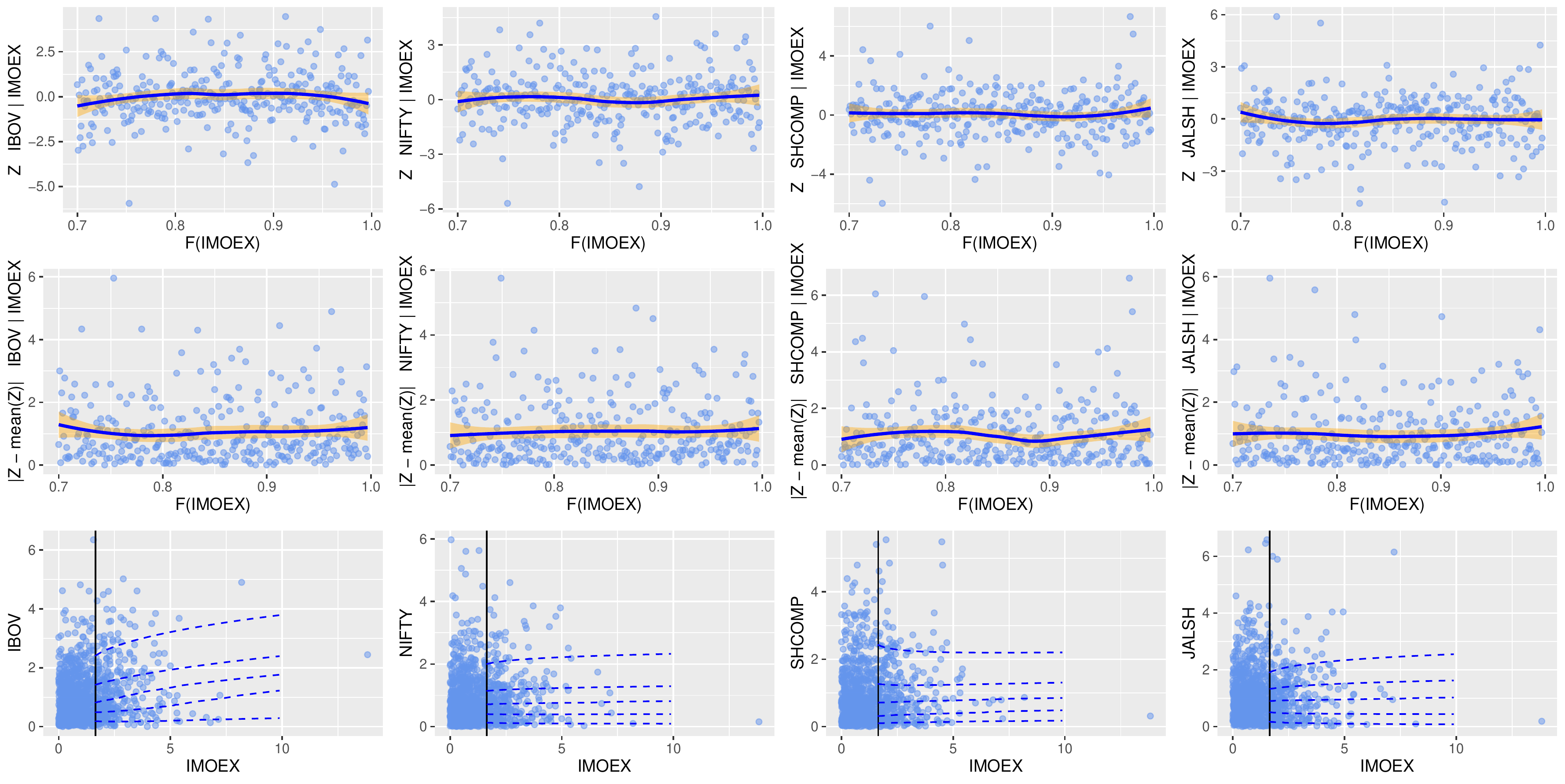}
		\caption[Dependence model diagnostics: conditioning on the IMOEX variable]{Dependence model diagnostics: conditioning on the IMOEX variable.}
		\label{DepQuaq2}
	\end{figure}
	
	\begin{figure}[H]
		\centering
		\includegraphics[width=10.5 cm]{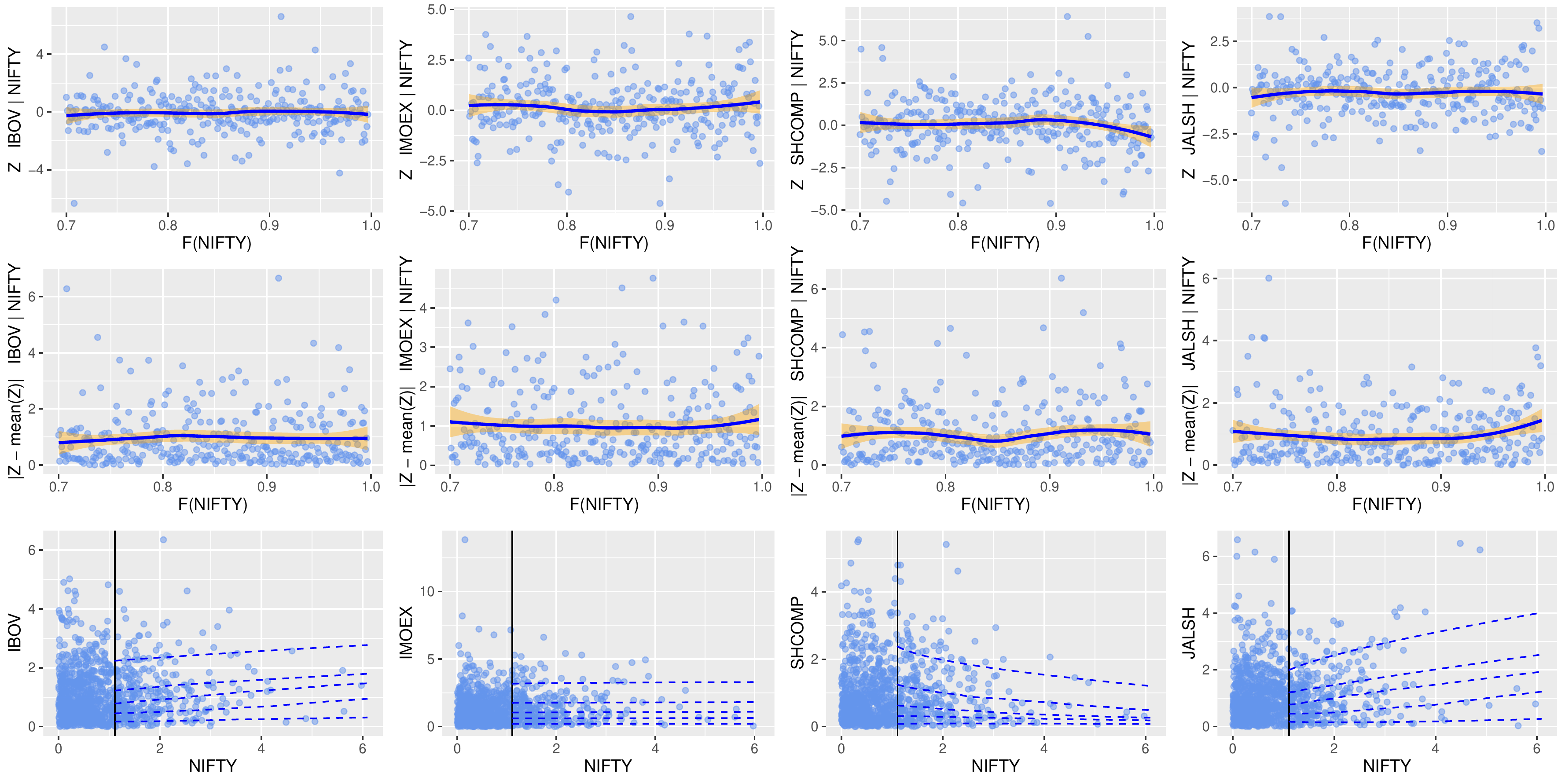}
		\caption[Dependence model diagnostics: conditioning on the NIFTY variable]{Dependence model diagnostics: conditioning on the NIFTY variable.}
		\label{DepQuaq3}
		%\end{figure}
		\begin{flushleft}
		\end{flushleft}
		%\begin{figure}
		\centering
		\includegraphics[width=10.5 cm]{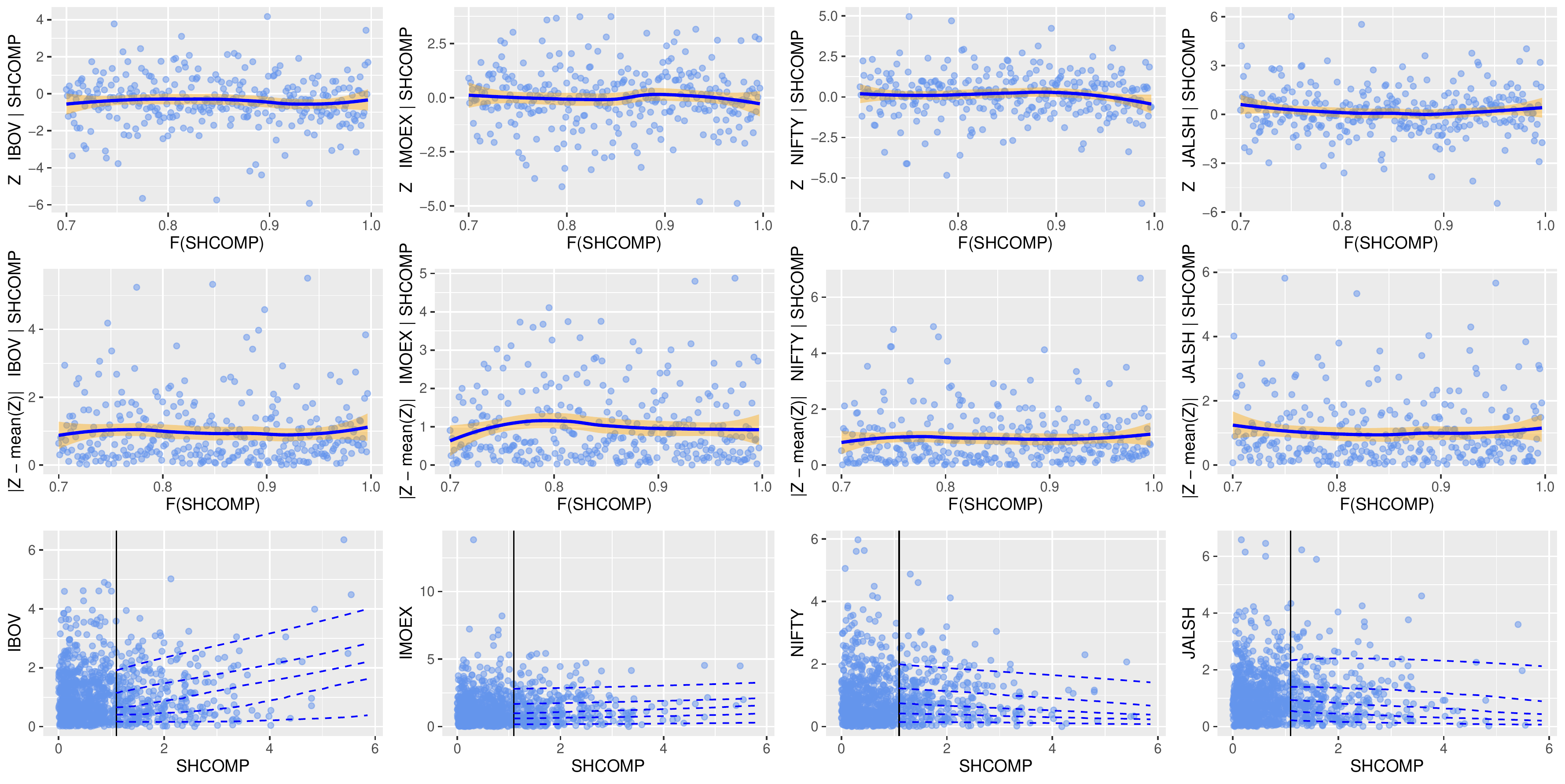}
		\caption[Dependence model diagnostics: conditioning on the SHCOMP variable]{Dependence model diagnostics: conditioning on the SHCOMP variable.}
		\label{DepQuaq4}
	\end{figure}
	
	\begin{figure}[H]
		\centering
		\includegraphics[width=10.5 cm]{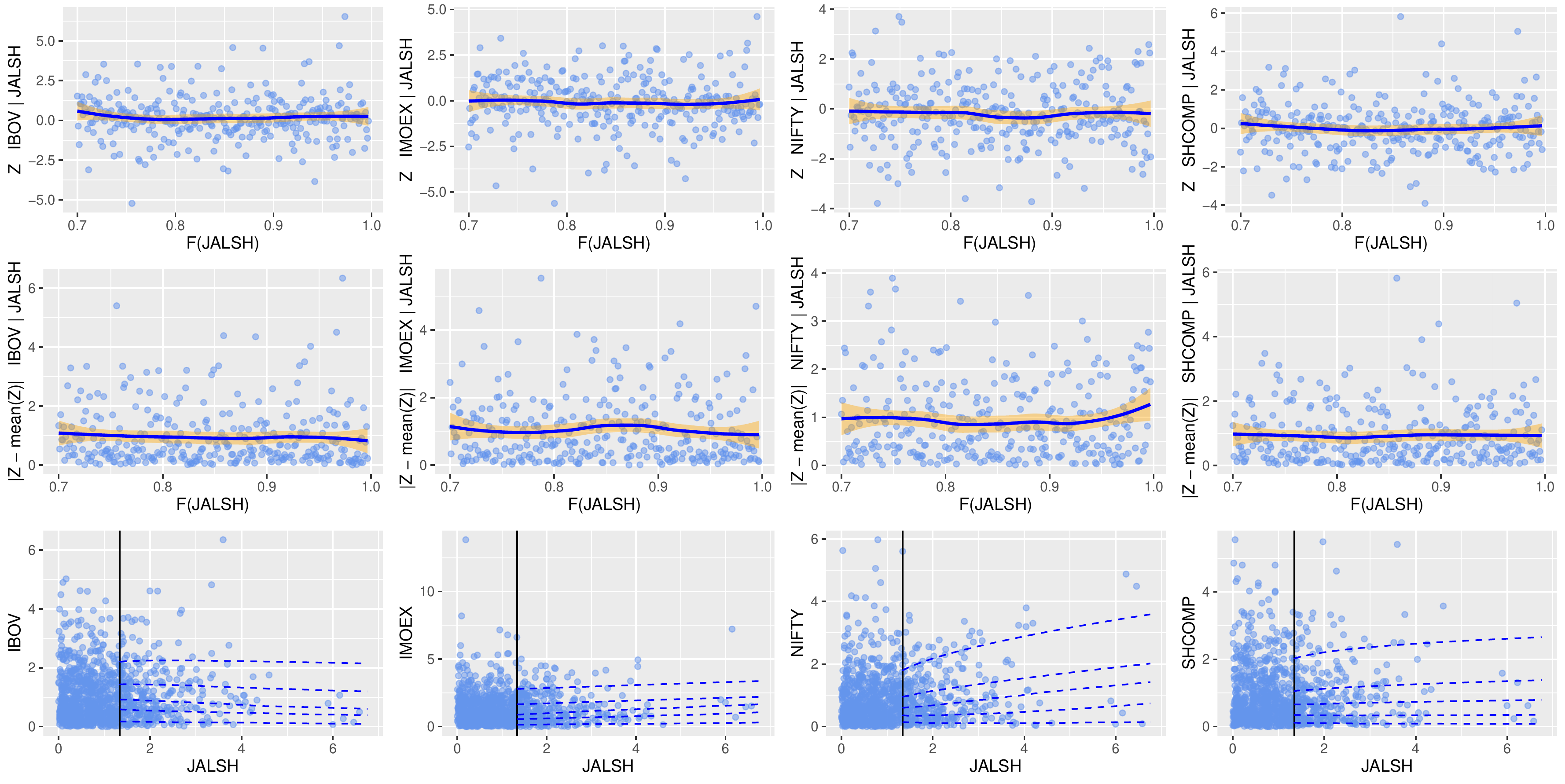}
		\caption[Dependence model diagnostics: conditioning on the JALSH variable]{Dependence model diagnostics: conditioning on the JALSH variable.}
		\label{DepQuaq5}
	\end{figure}

	%\subsection*{Appendix A3: Prediction plots}

%	
%	%%%%%%%%%%%%%%%%%%%%%%%%%%%%%%%%%%%%%%%%%%
%\end{paracol}
%%%%%%%%%%%%%%%%%%%%%%%%%%%%%%%%%%%%%%%%%%%

%\reftitle{References}


%\externalbibliography{yes}
%\bibliography{your_external_BibTeX_file}

%=====================================
% References, variant B: internal bibliography
%=====================================

%%%%%%%%%%%%%%%%%%%%%%%%%%%%%%%%%%%%%%%%
\end{document}